\documentclass[12pt]{article}
\usepackage{amsmath,amssymb}
\usepackage{graphicx}
\newcommand{\eqdef}{\stackrel{\text{def}}{=}}

\newcommand{\n}{\nonumber\\}
\newcommand{\bm}{\boldsymbol}
\newcommand{\ignore}[1]{}
\numberwithin{equation}{section}
\newcommand{\Romannumeral}[1]{\uppercase\expandafter{\romannumeral#1}}
\newcommand{\I}{\text{\Romannumeral{1}}}
\newcommand{\II}{\text{\Romannumeral{2}}}
\newcommand{\III}{\text{\Romannumeral{3}}}
\newcommand{\IV}{\text{\Romannumeral{4}}}
\newcommand{\V}{\text{\Romannumeral{5}}}
\newcommand{\VI}{\text{\Romannumeral{6}}}
\newtheorem{thm}{\bf Theorem}
\newtheorem{conj}{\bf Conjecture}

\allowdisplaybreaks[4]

\begin{document}

\baselineskip=20pt

\newfont{\elevenmib}{cmmib10 scaled\magstep1}
\newcommand{\preprint}{
    \begin{flushright}\normalsize \sf
     DPSU-19-2\\
   \end{flushright}}
\newcommand{\Title}[1]{{\baselineskip=26pt
   \begin{center} \Large \bf #1 \\ \ \\ \end{center}}}
\newcommand{\Author}{\begin{center}
   \large \bf Satoru Odake \end{center}}
\newcommand{\Address}{\begin{center}
     Faculty of Science, Shinshu University,\\
     Matsumoto 390-8621, Japan
   \end{center}}
\newcommand{\Accepted}[1]{\begin{center}
   {\large \sf #1}\\ \vspace{1mm}{\small \sf Accepted for Publication}
   \end{center}}

\preprint
\thispagestyle{empty}

\Title{Recurrence Relations of\\
the Multi-Indexed Orthogonal Polynomials $\VI$ :\\
Meixner-Pollaczek and continuous Hahn types}

\Author

\Address
\vspace{1cm}

\begin{abstract}
In previous papers, we discussed the recurrence relations of the
multi-indexed orthogonal polynomials of the Laguerre, Jacobi, Wilson,
Askey-Wilson, Racah and $q$-Racah types.
In this paper we explore those of the Meixner-Pollaczek and continuous
Hahn types.
For the $M$-indexed Meixner-Pollaczek and continuous Hahn polynomials,
we present $3+2M$ term recurrence relations with variable dependent
coefficients and $1+2L$ term ($L\geq M+1$) recurrence relations with
constant coefficients.
Based on the latter, the generalized closure relations and the
creation/annihilation operators of the quantum mechanical systems described
by the multi-indexed Meixner-Pollaczek and continuous Hahn polynomials
are obtained.
\end{abstract}

\section{Introduction}
\label{intro}

Orthogonal polynomials in the Askey scheme of the (basic) hypergeometric
orthogonal polynomials \cite{kls}, which satisfy second order differential
or difference equations, can be successfully studied in the quantum mechanical
formulation: ordinary quantum mechanics (oQM), discrete quantum mechanics
with pure imaginary shifts (idQM) \cite{os13}--\cite{os24} and discrete
quantum mechanics with real shifts (rdQM) \cite{os12}--\cite{os34}.
By deforming the exactly solvable quantum systems described by the orthogonal
polynomials in the Askey scheme, new types of orthogonal polynomials are
obtained \cite{gkm08}--\cite{dualmiopqR}.
They are called exceptional or multi-indexed orthogonal polynomials
$\{\mathcal{P}_n(\eta)|n\in\mathbb{Z}_{\geq 0}\}$ (the range of $n$ is finite
for finite rdQM systems).
They satisfy second order differential or difference equations and
form a complete set of orthogonal basis in an appropriate Hilbert space
in spite of missing degrees.
This degree missing is a characteristic feature of them, and thereby the
Bochner's theorem and its generalizations \cite{bochner,szego} are avoided.
We distinguish the following two cases;
the set of missing degrees $\mathcal{I}=\mathbb{Z}_{\geq 0}\backslash
\{\deg\mathcal{P}_n|n\in\mathbb{Z}_{\geq 0}\}$ is
case-(1): $\mathcal{I}=\{0,1,\ldots,\ell-1\}$, or
case-(2): $\mathcal{I}\neq\{0,1,\ldots,\ell-1\}$, where $\ell$ is a positive
integer. The situation of case-(1) is called stable in \cite{gkm11}.

Ordinary orthogonal polynomials in one variable are characterized by the
three term recurrence relations \cite{szego}.
Since the multi-indexed orthogonal polynomials are not ordinary orthogonal
polynomials, they do not satisfy the three term recurrence relations.
Instead of the three term recurrence relations, they satisfy some recurrence
relations with more terms \cite{stz10}--\cite{rrmiop5}.

In previous papers \cite{rrmiop,rrmiop2,rrmiop3,rrmiop4,rrmiop5},
the recurrence relations for the case-(1) multi-indexed orthogonal polynomials
(Laguerre (L) and Jacobi (J) types in oQM \cite{os25},
Wilson (W) and Askey-Wilson (AW) types in idQM \cite{os27},
Racah (R) and $q$-Racah ($q$R) types in rdQM \cite{os26}) were studied.
There are two kinds of recurrence relations:
with variable dependent coefficients and with constant coefficients.

Recently, based on the idQM systems whose physical range of the coordinate
is the whole real line, the case-(1) multi-indexed orthogonal polynomials of
Meixner-Pollaczek (MP) and continuous Hahn (cH) types are constructed
\cite{idQMcH}.
We remark that the MP and cH polynomials reduce to the Hermite (H) polynomial
in certain limits but there are no case-(1) multi-indexed Hermite orthogonal
polynomials.

In this paper we explore the recurrence relations for the case-(1)
multi-indexed orthogonal polynomials of MP and cH types.
By similar methods used in W and AW cases, we derive two kinds of recurrence
relations.
The recurrence relations with constant coefficients are closely related to
the generalized closure relations \cite{rrmiop4}.
The generalized closure relations provide the exact Heisenberg operator
solution of a certain operator, from which the creation and annihilation
operators of the system are obtained.
We present the creation and annihilation operators of the deformed MP and cH
systems.

The deformed Hamiltonian $\mathcal{H}_{\mathcal{D}}$ is determined by the
denominator polynomial $\Xi_{\mathcal{D}}(\eta)$, where $\mathcal{D}$ is an
index set of the virtual state wavefunctions used in $M$-step Darboux
transformations.
The degree of $\Xi_{\mathcal{D}}(\eta)$ is given by $\ell_{\mathcal{D}}$
\eqref{lD} and there is no restriction on $\ell_{\mathcal{D}}$ for L, J, W
and AW cases.
However, for MP and cH cases, the degree $\ell_{\mathcal{D}}$ must be even
in order that the Hamiltonian $\mathcal{H}_{\mathcal{D}}$ is hermitian.
The multi-indexed MP and cH orthogonal polynomials are constructed for even
$\ell_{\mathcal{D}}$ (and some conditions) \cite{idQMcH}.
In this paper we define the multi-indexed MP and cH polynomials for any index
set $\mathcal{D}$, namely $\ell_{\mathcal{D}}$ may be odd and they may not be
orthogonal polynomials.
We conjecture that the recurrence relations for the multi-indexed MP and cH
polynomials hold for non-orthogonal case.

This paper is organized as follows.
In section \ref{sec:miop} the case-(1) multi-indexed MP and cH polynomials
are recapitulated.
The definitions of the multi-indexed MP and cH polynomials are extended to
non-orthogonal case.
In section \ref{sec:rr} the recurrence relations of the case-(1) multi-indexed
MP and cH orthogonal polynomials are derived.
The recurrence relations with variable dependent coefficients are presented
in section \ref{sec:rr_var}, and those with constant coefficients in section
\ref{sec:rr_const}.
Some explicit examples are presented in section \ref{sec:ex}.
In section \ref{sec:gcr} the generalized closure relations and the
creation/annihilation operators are presented.
Section \ref{sec:summary} is for a summary and comments.
In Appendix \ref{app:idQM} formulation of idQM and deformed systems are
recapitulated.
Some properties of the multi-indexed MP polynomials are presented in Appendix
\ref{app:MP:prop}, and those of the multi-indexed cH polynomials are presented
in Appendix \ref{app:cH:prop}.
In Appendix \ref{app:ex} more examples for section \ref{sec:rr} are presented,
which correspond to non-orthogonal case.

\section{Multi-indexed Meixner-Pollaczek and Continuous\\ Hahn polynomials}
\label{sec:miop}

In this section we recapitulate the case-(1) multi-indexed Meixner-Pollaczek
and continuous Hahn polynomials \cite{idQMcH}.
Generalizing the result of \cite{idQMcH}, we first define the multi-indexed
MP and cH polynomials for any $\mathcal{D}$, and then consider the condition
that they become orthogonal polynomials.

The case-(1) multi-indexed orthogonal polynomials
\cite{os25,os27,os26,os35,idQMcH} and those of case-(2)
\cite{os15,gos,os22,os29,os30,casoidrdqm} are constructed based on the
quantum mechanical formulations \cite{os24}.
For the Meixner-Pollaczek (MP) and Continuous Hahn (cH) polynomials,
we use the discrete quantum mechanics with pure imaginary shifts (idQM)
\cite{os13,os24}.
The formulation of idQM is presented in Appendix \ref{app:idQM} and we follow
the notation there.
For MP and cH cases, the lower bound $x_1$, upper bound $x_2$, the
parameter $\gamma$, the sinusoidal coordinate $\eta(x)$ and the auxiliary
function $\varphi(x)$ are
\begin{equation}
  x_1=-\infty,\quad x_2=\infty,\quad\gamma=1,\quad\eta(x)=x,\quad
  \varphi(x)=1.
  \label{x1x2gamma}
\end{equation}
Namely, the physical range of the coordinate $x$ is the whole real line.
It is not necessary to distinguish $\check{P}_n$ and $P_n$ since $\eta(x)=x$,
but we will use both notations to compare with other cases in \cite{os27}.

\subsection{Meixner-Pollaczek and continuous Hahn polynomials}
\label{sec:MPcH}

First we take a set of parameters $\bm{\lambda}$ as follows:
\begin{equation}
  \text{MP :}\ \ \bm{\lambda}=(a,\phi),\ \ a,\phi\in\mathbb{R},\qquad
  \text{cH :}\ \ \bm{\lambda}=(a_1,a_2),\ \ a_1,a_2\in\mathbb{C}.
  \label{lambda}
\end{equation}
The fundamental data are the following \cite{os13}: ($n\in\mathbb{Z}_{\geq 0}$)
\begin{align}
  &V(x;\bm{\lambda})=\left\{
  \begin{array}{ll}
  e^{i(\frac{\pi}{2}-\phi)}(a+ix)&:\text{MP}\\
  (a_1+ix)(a_2+ix)&:\text{cH}
  \end{array}\right.,
  \label{Vform}\\
  &\mathcal{E}_n(\bm{\lambda})=\left\{
  \begin{array}{ll}
  2n\sin\phi&:\text{MP}\\
  n(n+b_1-1)&:\text{cH}
  \end{array}\right.,\ \ b_1\eqdef a_1+a_2+a_1^*+a_2^*,
  \label{En}\\
  &\check{P}_n(x;\bm{\lambda})=P_n\bigl(\eta(x);\bm{\lambda}\bigr)=\left\{
  \begin{array}{ll}
  P^{(a)}_n\bigl(\eta(x);\phi\bigr)&:\text{MP}\\[2pt]
  p_n\bigl(\eta(x);a_1,a_2,a_1^*,a_2^*\bigr)&:\text{cH}
  \end{array}\right.
  \label{Pn}\\
  &\phantom{\check{P}_n(x;\bm{\lambda})}=\left\{
  \begin{array}{ll}
  {\displaystyle\frac{(2a)_n}{n!}e^{in\phi}
  {}_2F_1\Bigl(\genfrac{}{}{0pt}{}{-n,\,a+ix}
  {2a}\Bigm|1-e^{-2i\phi}\Bigr)}&:\text{MP}\\[10pt]
  {\displaystyle i^n\frac{(a_1+a_1^*,a_1+a_2^*)_n}{n!}
  {}_3F_2\Bigl(\genfrac{}{}{0pt}{}{-n,\,n+b_1-1,\,a_1+ix}
  {a_1+a_1^*,\,a_1+a_2^*}\Bigm|1\Bigr)}&:\text{cH}
  \end{array}\right.\\
  &\phantom{\check{P}_n(x;\bm{\lambda})}=
  c_n(\bm{\lambda})\eta(x)^n+(\text{lower order terms}),\quad
  c_n(\bm{\lambda})=\left\{
  \begin{array}{ll}
  \frac{1}{n!}(2\sin\phi)^n&:\text{MP}\\[4pt]
  \frac{1}{n!}(n+b_1-1)_n&:\text{cH}
  \end{array}\right.,
  \label{cn}\\
  &\bm{\delta}=\left\{
  \begin{array}{ll}
  (\tfrac12,0)&:\text{MP}\\[2pt]
  (\tfrac12,\tfrac12)&:\text{cH}
  \end{array}\right.,\ \ \kappa=1,
  \ \ f_n(\bm{\lambda})=\left\{
  \begin{array}{ll}
  2\sin\phi&:\text{MP}\\
  n+b_1-1&:\text{cH}
  \end{array}\right.,
  \ \ b_{n-1}(\bm{\lambda})=n.
\end{align}
(Although the notation $b_1$ conflicts with $b_{n-1}(\bm{\lambda})$,
we think this does not cause any confusion.)
Here $P^{(a)}_n(\eta;\phi)$ and $p_n(\eta;a_1,a_2,a_3,a_4)$ in \eqref{Pn} are
the Meixner-Pollaczek and continuous Hahn polynomials of degree $n$ in $\eta$,
respectively \cite{kls}.
Note that $\check{P}^*_n(x;\bm{\lambda})=\check{P}_n(x;\bm{\lambda})$.
The Meixner-Pollaczek polynomial with $\phi=\frac{\pi}{2}$ has a definite
parity,
\begin{equation}
  \text{MP}:\ \ \check{P}_n(-x;\bm{\lambda})=(-1)^n\check{P}_n(x;\bm{\lambda})
  \ \ \text{for $\bm{\lambda}=(a,\tfrac{\pi}{2})$}.
  \label{MPparity}
\end{equation}

The polynomials $\check{P}_n(x)$ satisfy the forward and backward shift
relations \eqref{FP=,BP=}, hence the second order difference equation
\eqref{HtP=EP}.

Next let us restrict a set of parameters $\bm{\lambda}$ as follows:
\begin{equation}
  \text{MP :}\ \ a>0,\ \ 0<\phi<\pi,\qquad
  \text{cH :}\ \ \text{Re}\,a_i>0\ \ (i=1,2).
  \label{pararange}
\end{equation}
Then the Hamiltonian of idQM system is well-defined and hermitian.
Additional fundamental data are the following \cite{os13}:
\begin{align}
  &\phi_n(x;\bm{\lambda})
  =\phi_0(x;\bm{\lambda})\check{P}_n(x;\bm{\lambda}),
  \label{factphin}\\
  &\phi_0(x;\bm{\lambda})=\left\{
  \begin{array}{ll}
  e^{(\phi-\frac{\pi}{2})x}\sqrt{\Gamma(a+ix)\Gamma(a-ix)}&:\text{MP}\\[4pt]
  \sqrt{\Gamma(a_1+ix)\Gamma(a_2+ix)\Gamma(a_1^*-ix)\Gamma(a_2^*-ix)}
  &:\text{cH}
  \end{array}\right.,
  \label{phi0}\\
  &h_n(\bm{\lambda})=\left\{
  \begin{array}{ll}
  {\displaystyle 2\pi\frac{\Gamma(n+2a)}{n!\,(2\sin\phi)^{2a}}}
  &:\text{MP}\\[10pt]
  {\displaystyle 2\pi\frac{\prod_{j,k=1}^2\Gamma(n+a_j+a_k^*)}
  {n!\,(2n+b_1-1)\Gamma(n+b_1-1)}}&:\text{cH}
  \end{array}\right..
  \label{hn}
\end{align}
Note that $\phi^*_0(x;\bm{\lambda})=\phi_0(x;\bm{\lambda})$.
The eigenfunctions $\phi_n(x)$ are orthogonal each other \eqref{(phin,phim)},
which gives the orthogonality relation of $\check{P}_n(x)$ \eqref{orthocPn}.

The MP and cH idQM systems have shape invariance property \eqref{shapeinv},
which gives the forward and backward shift relations \eqref{FP=,BP=}.
The second order difference equation \eqref{HtP=EP} is a rewrite of the
Schr\"odinger equation \eqref{Hphin=}.

\subsection{Multi-indexed Meixner-Pollaczek and continuous Hahn polynomials}
\label{sec:miopMPcH}

The case-(1) multi-indexed Meixner-Pollaczek and continuous Hahn polynomials
were constructed by deforming the idQM systems in \S\,\ref{sec:MPcH}
\cite{idQMcH}.
The index set $\mathcal{D}=\{d_1,\ldots,d_M\}$
($d_j$: mutually distinct) labels the virtual state wavefunctions used in the
$M$-step Darboux transformations.
For cH system, there are two types of virtual states (type $\I$ and $\II$)
and $\mathcal{D}$ is
$\mathcal{D}=\{d_1,\ldots,d_M\}=\{d^{\I}_1,\ldots,d^{\I}_{M_{\I}},
d^{\II}_1,\ldots,d^{\II}_{M_{\II}}\}$ ($M=M_{\I}+M_{\II}$,
$d^{\I}_j$ : mutually distinct,
$d^{\II}_j$ : mutually distinct).

First we define multi-indexed MP and cH polynomials for any $\bm{\lambda}$
and $\mathcal{D}$, which may not be orthogonal polynomials. Then we present
a sufficient condition for them to become orthogonal polynomials.

\subsubsection{definitions (for any $\bm{\lambda}$ and $\mathcal{D}$)}
\label{sec:miopdef}

A set of parameters $\bm{\lambda}$ is taken as \eqref{lambda} and
$d_j$'s are non-negative integers.
The twist operations $\mathfrak{t}$ and constants $\tilde{\bm{\delta}}$
are defined by
\begin{align}
  \text{MP :}&\ \ \mathfrak{t}(\bm{\lambda})\eqdef(1-a,\phi),\quad
  \tilde{\bm{\delta}}\eqdef(-\tfrac12,0),\\
  \text{cH :}&\ \ \ \text{type $\I$}:\ \ \ \mathfrak{t}^{\I}(\bm{\lambda})
  \eqdef(1-a_1^*,a_2),\quad\,
  \tilde{\bm{\delta}}^{\I}\eqdef(-\tfrac12,\tfrac12),\n
  &\ \ \text{type $\II$}:\ \ \mathfrak{t}^{\II}(\bm{\lambda})
  \eqdef(a_1,1-a_2^*),\quad
  \tilde{\bm{\delta}}^{\II}\eqdef(\tfrac12,-\tfrac12).
\end{align}
For cH case, corresponding to the type $\I$ and type $\II$, we add superscripts
$\I$ and $\II$.
The virtual state energies $\tilde{\mathcal{E}}_{\text{v}}(\bm{\lambda})$ and
virtual state polynomials $\xi_{\text{v}}(\eta;\bm{\lambda})$ are defined by
\begin{align}
  \text{MP :}&\ \ \tilde{\mathcal{E}}_{\text{v}}(\bm{\lambda})
  \eqdef-2(2a-\text{v}-1)\sin\phi,\n
  &\ \ \check{\xi}_{\text{v}}(x;\bm{\lambda})\eqdef
  \xi_{\text{v}}\bigl(\eta(x);\bm{\lambda}\bigr)\eqdef
  \check{P}_{\text{v}}\bigl(x;\mathfrak{t}(\bm{\lambda})\bigr)
  =P_{\text{v}}\bigl(\eta(x);\mathfrak{t}(\bm{\lambda})\bigr),\\
  \text{cH :}&\ \ \tilde{\mathcal{E}}^{\I}_{\text{v}}(\bm{\lambda})
  \eqdef-(a_1+a_1^*-\text{v}-1)(a_2+a_2^*+\text{v}),\n
  &\ \ \tilde{\mathcal{E}}^{\II}_{\text{v}}(\bm{\lambda})
  \eqdef-(a_2+a_2^*-\text{v}-1)(a_1+a_1^*+\text{v}),\n
  &\ \ \check{\xi}^{\I}_{\text{v}}(x;\bm{\lambda})\eqdef
  \xi^{\I}_{\text{v}}\bigl(\eta(x);\bm{\lambda}\bigr)\eqdef
  \check{P}_{\text{v}}\bigl(x;\mathfrak{t}^{\I}(\bm{\lambda})\bigr)
  =P_{\text{v}}\bigl(\eta(x);\mathfrak{t}^{\I}(\bm{\lambda})\bigr),\\
  &\ \ \check{\xi}^{\II}_{\text{v}}(x;\bm{\lambda})\eqdef
  \xi^{\II}_{\text{v}}\bigl(\eta(x);\bm{\lambda}\bigr)\eqdef
  \check{P}_{\text{v}}\bigl(x;\mathfrak{t}^{\II}(\bm{\lambda})\bigr)
  =P_{\text{v}}\bigl(\eta(x);\mathfrak{t}^{\II}(\bm{\lambda})\bigr).
  \nonumber
\end{align}
The virtual state polynomials $\xi_{\text{v}}(\eta;\bm{\lambda})$ are
polynomials of degree $\text{v}$ in $\eta$ and satisfy
$\widetilde{\mathcal{H}}(\bm{\lambda})\check{\xi}_{\text{v}}(x;\bm{\lambda})$
$=\tilde{\mathcal{E}}_{\text{v}}(\bm{\lambda})
\check{\xi}_{\text{v}}(x;\bm{\lambda})$.
Note that $\check{\xi}^*_{\text{v}}(x;\bm{\lambda})
=\check{\xi}_{\text{v}}(x;\bm{\lambda})$.
The functions $r_j(x^{(M)}_j;\bm{\lambda},M)$ ($j=1,2,\ldots,M$) are defined
by
\begin{align}
  \text{MP :}&\ \ r_j(x^{(M)}_j;\bm{\lambda},M)
  \eqdef(-1)^{j-1}i^{1-M}(a-\tfrac{M-1}{2}+ix)_{j-1}
  (a-\tfrac{M-1}{2}-ix)_{M-j},\\
  \text{cH :}&\ \ r^{\I}_j(x^{(M)}_j;\bm{\lambda},M)
  \eqdef(-1)^{j-1}i^{1-M}(a_1-\tfrac{M-1}{2}+ix)_{j-1}
  (a_1^*-\tfrac{M-1}{2}-ix)_{M-j},\n
  &\ \,r^{\II}_j(x^{(M)}_j;\bm{\lambda},M)
  \eqdef(-1)^{j-1}i^{1-M}(a_2-\tfrac{M-1}{2}+ix)_{j-1}
  (a_2^*-\tfrac{M-1}{2}-ix)_{M-j},
\end{align}
where $x_j^{(n)}\eqdef x+i(\tfrac{n+1}{2}-j)\gamma$.
The auxiliary function $\varphi_M(x)$ introduced in \cite{gos} is
$\varphi_M(x)=1$ in the present case.

Let us define the denominator polynomial
$\Xi_{\mathcal{D}}(\eta;\bm{\lambda})$ and the multi-indexed polynomial
$P_{\mathcal{D},n}(\eta;\bm{\lambda})$ :
\begin{equation}
  \check{\Xi}_{\mathcal{D}}(x;\bm{\lambda})\eqdef
  \Xi_{\mathcal{D}}\bigl(\eta(x);\bm{\lambda}\bigr),\quad
  \check{P}_{\mathcal{D},n}(x;\bm{\lambda})\eqdef
  P_{\mathcal{D},n}\bigl(\eta(x);\bm{\lambda}\bigr).
  \label{XiP_poly_la}
\end{equation}
Here $\check{\Xi}_{\mathcal{D}}(x;\bm{\lambda})$ and
$\check{P}_{\mathcal{D},n}(x;\bm{\lambda})$ are given by determinants
as follows. For MP, they are
\begin{align}
  &i^{\frac12M(M-1)}\det\bigl(\check{\xi}_{d_k}(x^{(M)}_j;\bm{\lambda})
  \bigr)_{1\leq j,k\leq M}
  =\varphi_M(x)\check{\Xi}_{\mathcal{D}}(x;\bm{\lambda}),
  \label{MP:cXiDdef}\\
  &i^{\frac12M(M+1)}\left|
  \begin{array}{cccc}
  \check{\xi}_{d_1}(x^{(M+1)}_1;\bm{\lambda})&\cdots&
  \check{\xi}_{d_M}(x^{(M+1)}_1;\bm{\lambda})
  &r_1(x^{(M+1)}_1;\bm{\lambda},M+1)\check{P}_n(x^{_(M+1)}_1;\bm{\lambda})\\
  \check{\xi}_{d_1}(x^{(M+1)}_2;\bm{\lambda})&\cdots&
  \check{\xi}_{d_M}(x^{(M+1)}_2;\bm{\lambda})
  &r_2(x^{(M+1)}_2;\bm{\lambda},M+1)\check{P}_n(x^{(M+1)}_2;\bm{\lambda})\\
  \vdots&\cdots&\vdots&\vdots\\
  \check{\xi}_{d_1}(x^{(M+1)}_{M+1};\bm{\lambda})&\cdots&
  \check{\xi}_{d_M}(x^{(M+1)}_{M+1};\bm{\lambda})
  &r_{M+1}(x^{(M+1)}_{M+1};\bm{\lambda},M+1)
  \check{P}_n(x^{(M+1)}_{M+1};\bm{\lambda})\\
  \end{array}\right|\n
  &=\varphi_{M+1}(x)\check{P}_{\mathcal{D},n}(x;\bm{\lambda}).
  \label{MP:cPDndef}
\end{align}
For cH, they are
\begin{align}
  &i^{\frac12M(M-1)}\left|
  \begin{array}{llllll}
  \vec{X}^{(M)}_{d^{\I}_1}&\cdots&\vec{X}^{(M)}_{d^{\I}_{M_{\I}}}&
  \vec{Y}^{(M)}_{d^{\II}_1}&\cdots&\vec{Y}^{(M)}_{d^{\II}_{M_{\II}}}\\
  \end{array}\right|
  =\varphi_M(x)\check{\Xi}_{\mathcal{D}}(x;\bm{\lambda})\times A,
  \label{cXiDdef}\\
  &i^{\frac12M(M+1)}\left|
  \begin{array}{lllllll}
  \vec{X}^{(M+1)}_{d^{\I}_1}&\cdots&\vec{X}^{(M+1)}_{d^{\I}_{M_{\I}}}&
  \vec{Y}^{(M+1)}_{d^{\II}_1}&\cdots&\vec{Y}^{(M+1)}_{d^{\II}_{M_{\II}}}&
  \vec{Z}^{(M+1)}_n\\
  \end{array}\right|\n
  &=\varphi_{M+1}(x)\check{P}_{\mathcal{D},n}(x;\bm{\lambda})\times B,
  \label{cPDndef}
\end{align}
where $A$ and $B$ are
\begin{align}
  &A=\prod_{j=1}^{M_{\I}-1}
  (a_2-\tfrac{M-1}{2}+ix,a_2^*-\tfrac{M-1}{2}-ix)_j
  \cdot\prod_{j=1}^{M_{\II}-1}
  (a_1-\tfrac{M-1}{2}+ix,a_1^*-\tfrac{M-1}{2}-ix)_j,
  \label{cXiDA}\\
  &B=\prod_{j=1}^{M_{\I}}
  (a_2-\tfrac{M}{2}+ix,a_2^*-\tfrac{M}{2}-ix)_j
  \cdot\prod_{j=1}^{M_{\II}}
  (a_1-\tfrac{M}{2}+ix,a_1^*-\tfrac{M}{2}-ix)_j,
  \label{cPDnB}
\end{align}
and $\vec{X}^{(M)}_{\text{v}}$, $\vec{Y}^{(M)}_{\text{v}}$ and
$\vec{Z}^{(M)}_{\text{v}}$ are
\begin{align}
  &\bigl(\vec{X}^{(M)}_{\text{v}}\bigr)_j
  =r^{\II}_j(x^{(M)}_j;\bm{\lambda},M)
  \check{\xi}^{\I}_{\text{v}}(x^{(M)}_j;\bm{\lambda}),\qquad
  (j=1,2,\ldots,M),\n
  &\bigl(\vec{Y}^{(M)}_{\text{v}}\bigr)_j
  =r^{\I}_j(x^{(M)}_j;\bm{\lambda},M)
  \check{\xi}^{\II}_{\text{v}}(x^{(M)}_j;\bm{\lambda}),\n
  &\bigl(\vec{Z}^{(M)}_n\bigr)_j
  =r^{\II}_j(x^{(M)}_j;\bm{\lambda},M)r^{\I}_j(x^{(M)}_j;\bm{\lambda},M)
  \check{P}_n(x^{(M)}_j;\bm{\lambda}).
\end{align}
(For the cases of type $\I$ only ($M_{\II}=0$) or type $\II$ only ($M_{\I}=0$),
the expressions \eqref{cXiDdef} and \eqref{cPDndef} are rewritten as
\eqref{MP:cXiDdef} and \eqref{MP:cPDndef}, see \cite{idQMcH}.)
The denominator polynomial $\Xi_{\mathcal{D}}(\eta;\bm{\lambda})$ and
the multi-indexed polynomial $P_{\mathcal{D},n}(\eta;\bm{\lambda})$ are
polynomials in $\eta$ and their degrees are $\ell_{\mathcal{D}}$ and
$\ell_{\mathcal{D}}+n$, respectively (we assume
$c_{\mathcal{D}}^{\Xi}(\bm{\lambda})\neq 0$ and
$c_{\mathcal{D},n}^{P}(\bm{\lambda})\neq 0$,
see \eqref{MP:cXiD}--\eqref{MP:cPDn} and \eqref{cXiD}--\eqref{cPDn}).
Here $\ell_{\mathcal{D}}$ is given by \eqref{lD}.
Note that $\check{\Xi}^*_{\mathcal{D}}(x;\bm{\lambda})
=\check{\Xi}_{\mathcal{D}}(x;\bm{\lambda})$ and
$\check{P}^*_{\mathcal{D},n}(x;\bm{\lambda})
=\check{P}_{\mathcal{D},n}(x;\bm{\lambda})$.
Under the permutation of $d_j$'s, $\check{\Xi}_{\mathcal{D}}(x)$ and
$\check{P}_{\mathcal{D},n}(x)$ change their signs.
The parity property of the Meixner-Pollaczek polynomial \eqref{MPparity}
is inherited by the multi-indexed polynomials of even $\ell_{\mathcal{D}}$,
\begin{equation}
  \text{MP}:\ \ \check{P}_{\mathcal{D},n}(-x;\bm{\lambda})
  =(-1)^n\check{P}_{\mathcal{D},n}(x;\bm{\lambda})
  \ \ \text{for $\bm{\lambda}=(a,\tfrac{\pi}{2})$ and even
  $\ell_{\mathcal{D}}$}.
  \label{MPmiopparity}
\end{equation}

The deformed potential functions $V_{\mathcal{D}}(x;\bm{\lambda})$ are
defined by \eqref{VD} and \eqref{la'}.
The multi-indexed polynomials $\check{P}_{\mathcal{D},n}(x;\bm{\lambda})$
satisfy the forward and backward shift relations \eqref{FDPDn=,BDPDn=},
hence the second order difference equation \eqref{tHPDn=}.

\subsubsection{orthogonal polynomials}
\label{sec:miopcond}

The deformed idQM systems should be well-defined, namely the deformed
Hamiltonian $\mathcal{H}_{\mathcal{D}}(\bm{\lambda})$ should be hermitian.
We restrict $\bm{\lambda}$ as \eqref{pararange} and impose the following
conditions,
\begin{align}
  \text{MP :}&\ \ \max_j\{d_j\}<2a-1,\\
  \text{cH :}&\ \ \max_j\{d^{\I}_j\}<a_1+a_1^*-1,\quad
  \max_j\{d^{\II}_j\}<a_2+a_2^*-1,
\end{align}
under which the virtual state energies become negative,
\begin{align*}
  \text{MP :}&\ \ 
  \tilde{\mathcal{E}}_{\text{v}}(\bm{\lambda})<0
 \ \Leftrightarrow\ 2a>\text{v}+1,\n
  \text{cH :}&\ \ 
  \tilde{\mathcal{E}}^{\I}_{\text{v}}(\bm{\lambda})<0
  \ \Leftrightarrow\ a_1+a_1^*>\text{v}+1,\quad
  \tilde{\mathcal{E}}^{\II}_{\text{v}}(\bm{\lambda})<0
  \ \Leftrightarrow\ a_2+a_2^*>\text{v}+1.
\end{align*}
The deformed Hamiltonian $\mathcal{H}_{\mathcal{D}}(\bm{\lambda})$ is
hermitian, if the condition \eqref{nozero} is satisfied \cite{idQMcH}.
Since the rectangular domain $D_{\gamma}$ contains the whole real axis,
the degree of $\Xi_{\mathcal{D}}(\eta;\bm{\lambda})$,
$\ell_{\mathcal{D}}$ \eqref{lD}, should be even.
Although we have no analytical proof that there exists a range of parameters
$\bm{\lambda}$ satisfying the condition \eqref{nozero},
we can verify that there exists such a range of $\bm{\lambda}$ by numerical
calculation (for small $M$ and $d_j$).
We have observed various sufficient conditions for the parameter range
satisfying \eqref{nozero}, see \cite{idQMcH}.
In the rest of this section we assume that the condition \eqref{nozero} is
satisfied.

The eigenfunctions of the deformed Hamiltonian
$\mathcal{H}_{\mathcal{D}}(\bm{\lambda})$ have the form \eqref{phiDn}.
Note that
$\psi^*_{\mathcal{D}}(x;\bm{\lambda})=\psi_{\mathcal{D}}(x;\bm{\lambda})$.
The eigenfunctions $\phi_{\mathcal{D}\,n}(x)$ are orthogonal each other,
which gives the orthogonality relation of $\check{P}_{\mathcal{D},n}(x)$,
\eqref{orthocPDn}--\eqref{hDn}.
The multi-indexed orthogonal polynomial $P_{\mathcal{D},n}(\eta;\bm{\lambda})$
has $n$ zeros in the physical region $\eta\in\mathbb{R}$ ($\Leftrightarrow$
$\eta(x_1)<\eta<\eta(x_2)$), which interlace the $n+1$ zeros of
$P_{\mathcal{D},n+1}(\eta;\bm{\lambda})$ in the physical region,
and $\ell_{\mathcal{D}}$ zeros in the unphysical region
$\eta\in\mathbb{C}\backslash\mathbb{R}$.
These properties and \eqref{orthocPDn} can be verified by numerical calculation.

Since the deformed Hamiltonian $\mathcal{H}_{\mathcal{D}}(\bm{\lambda})$
\eqref{HD} is expressed in terms of the potential function
$V_{\mathcal{D}}(x;\bm{\lambda})$ \eqref{VD},
$\mathcal{H}_{\mathcal{D}}(\bm{\lambda})$
is determined by the denominator polynomial
$\check{\Xi}_{\mathcal{D}}(x;\bm{\lambda})$, whose normalization is irrelevant.
Under the permutation of $d_j$'s, the deformed Hamiltonian
$\mathcal{H}_{\mathcal D}$ is invariant.

The deformed MP and cH idQM systems have also shape invariance property
\eqref{shapeinvD}, which gives the forward and backward shift relations
\eqref{FDPDn=,BDPDn=}.
The second order difference equation \eqref{tHPDn=} is a rewrite of the
Schr\"odinger equation \eqref{HphiDn=}.
The properties \eqref{MP:dM=0} and \eqref{dIM1=0}--\eqref{dIIM2=0} are also
the consequences of the shape invariance.
By \eqref{MP:PD0=A.XiD} and \eqref{PD0=A.XiD}, the similar property holds
for the denominator polynomial $\check{\Xi}_{\mathcal{D}}(x;\bm{\lambda})$.
The $M$-step deformed system labeled by $\mathcal{D}$ with $0$ is equivalent
to the $(M-1)$-step deformed system labeled by $\mathcal{D}'$ with shifted
parameters $\bm{\lambda}+\tilde{\bm{\delta}}$.
For the multi-indexed W and AW polynomials, there are equivalence among
the index sets $\mathcal{D}$ \cite{equiv_miop}.
The multi-indexed MP and cH polynomials have also equivalence
in the same form as W and AW cases, which is derived from the properties
\eqref{MP:dM=0} and \eqref{dIM1=0}--\eqref{dIIM2=0}.

\section{Recurrence Relations}
\label{sec:rr}

In this section we present the recurrence relations of the case-(1)
multi-indexed Meixner-Pollaczek and continuous Hahn polynomials.
There are two types of recurrence relations: variable dependent coefficients
and constant coefficients.

The three term recurrence relations of the Meixner-Pollaczek and continuous
Hahn polynomials are \cite{kls}
\begin{equation}
  \eta P_n(\eta;\bm{\lambda})=A_n(\bm{\lambda})P_{n+1}(\eta;\bm{\lambda})
  +B_n(\bm{\lambda})P_n(\eta;\bm{\lambda})
  +C_n(\bm{\lambda})P_{n-1}(\eta;\bm{\lambda}),
  \label{3term}
\end{equation}
where $A_n$, $B_n$ and $C_n$ are
\begin{align}
  \text{MP :}&\ \ A_n(\bm{\lambda})=\frac{n+1}{2\sin\phi},\quad
  B_n(\bm{\lambda})=-(n+a)\cot\phi,\quad
  C_n(\bm{\lambda})=\frac{n+2a-1}{2\sin\phi},
  \label{MP:ABC}\\
  \text{cH :}&\ \ A_n(\bm{\lambda})
  =\frac{(n+1)(n+b_1-1)}{(2n+b_1-1)(2n+b_1)},\n
  &\ \ B_n(\bm{\lambda})=i\Bigl(a_1
  -\frac{(n+b_1-1)(n+a_1+a_1^*)(n+a_1+a_2^*)}{(2n+b_1-1)(2n+b_1)}\n
  &\phantom{\ \ B_n(\bm{\lambda})=i\Bigl(}
  +\frac{n(n+a_2+a_1^*-1)(n+a_2+a_2^*-1)}{(2n+b_1-2)(2n+b_1-1)}\Bigr),
  \label{ABC}\\
  &\ \ C_n(\bm{\lambda})
  =\frac{(n+a_1+a_1^*-1)(n+a_1+a_2^*-1)(n+a_2+a_1^*-1)(n+a_2+a_2^*-1)}
  {(2n+b_1-2)(2n+b_1-1)}.\nonumber
\end{align}
For simplicity of presentation, we set $P_n(\eta)=0$ for $n\in\mathbb{Z}_{<0}$
and define $A_n$, $B_n$ and $C_n$ for $n\in\mathbb{Z}_{<0}$ by
\eqref{MP:ABC}--\eqref{ABC}.
Then, \eqref{3term} hold for $n\in\mathbb{Z}$.
Similarly we set $P_{\mathcal{D},n}(\eta)=0$ for $n\in\mathbb{Z}_{<0}$.

\subsection{Recurrence relations with variable dependent coefficients}
\label{sec:rr_var}

We present $3+2M$ term recurrence relations with variable dependent
coefficients.
For the case-(1) multi-indexed Wilson and Askey-Wilson polynomials, such
recurrence relations are shown in \cite{rrmiop}.
Since the derivation can be applied in the present case without difficulty,
we present only the results here.

Let us define $\check{R}^{[s]}_{n,k}(x)$ ($n,k\in\mathbb{Z}$,
$s\in\mathbb{Z}_{\geq -1}$) as follows:
\begin{align}
  &\check{R}^{[s]}_{n,k}(x)=0\ \ (|k|>s+1\ \text{or}\ n+k<0),\quad
  \check{R}^{[-1]}_{n,0}(x)=1\ \ (n\geq 0),\n
  &\check{R}^{[s]}_{n,k}(x)
  =A_n\check{R}^{[s-1]}_{n+1,k-1}(x+i\tfrac{\gamma}{2})
  +\bigl(B_n-\eta(x-i\tfrac{s}{2}\gamma)\bigr)
  \check{R}^{[s-1]}_{n,k}(x+i\tfrac{\gamma}{2})
  \label{Rcdef}\\
  &\qquad\qquad
  +C_n\check{R}^{[s-1]}_{n-1,k+1}(x+i\tfrac{\gamma}{2})
  \ \ (s\geq 0).
  \nonumber
\end{align}
Here $A_n$, $B_n$ and $C_n$ are given by \eqref{MP:ABC}--\eqref{ABC}.
Note that $A_n$ ($n<-1$), $B_n$ ($n<0$) and $C_n$ ($n<0$) do not appear,
because $A_{-1}=0$ (and we regard $A_{-1}\times(\cdots)=0$).
For example, non-trivial $\check{R}^{[s]}_{n,k}(x)$ ($n+k\geq 0$)
for $s=0,1$ are
\begin{align*}
  s=0:\quad&\check{R}^{[0]}_{n,1}(x)=A_n,
  \ \ \check{R}^{[0]}_{n,0}(x)=B_n-\eta(x),
  \ \ \check{R}^{[0]}_{n,-1}(x)=C_n,\\
  s=1:\quad&\check{R}^{[1]}_{n,2}(x)=A_nA_{n+1},
  \ \ \check{R}^{[1]}_{n,1}(x)=A_n\bigl(B_n+B_{n+1}
  -\eta(x-i\tfrac{\gamma}{2})-\eta(x+i\tfrac{\gamma}{2})\bigr),\\
  &\check{R}^{[1]}_{n,0}(x)=A_nC_{n+1}+A_{n-1}C_n
  +\bigl(B_n-\eta(x-i\tfrac{\gamma}{2})\bigr)
  \bigl(B_n-\eta(x+i\tfrac{\gamma}{2})\bigr),\\
  &\check{R}^{[1]}_{n,-2}(x)=C_nC_{n-1},
  \ \ \check{R}^{[1]}_{n,-1}(x)=C_n\bigl(B_n+B_{n-1}
  -\eta(x-i\tfrac{\gamma}{2})-\eta(x+i\tfrac{\gamma}{2})\bigr).
\end{align*}
It is easy to see that $\check{R}^{[s]}_{n,k}(x)$ is a polynomial in
$x=\eta(x)$. We define $R^{[s]}_{n,k}(\eta)$ as follows:
\begin{equation}
  \check{R}^{[s]}_{n,k}(x)=R^{[s]}_{n,k}\bigl(\eta(x)\bigr)
  \ \ (|k|\leq s+1):
  \text{a polynomial of degree $s+1-|k|$ in $\eta(x)$}.
  \label{Rdef2}
\end{equation}
Note that $\check{R}^{[s]\,*}_{n,k}(x)=\check{R}^{[s]}_{n,k}(x)$.
Then we have the following result.
\begin{thm}
\label{thm:rr_var}
The multi-indexed Meixner-Pollaczek and continuous Hahn polynomials satisfy
the $3+2M$ term recurrence relations with variable dependent coefficients:
\begin{equation}
  \sum_{k=-M-1}^{M+1}R^{[M]}_{n,k}(\eta)P_{\mathcal{D},n+k}(\eta)=0,
  \label{RRP}
\end{equation}
which holds for $n\in\mathbb{Z}$.
\end{thm}
{\bf Remark 1}\,
The derivation in \cite{rrmiop} uses only the algebraic property of the
$M$-step Darboux transformations. Hence Theorem \ref{thm:rr_var} holds for
$P_{\mathcal{D},n}(\eta;\bm{\lambda})$ with any $\bm{\lambda}$ and
$\mathcal{D}$ (namely $P_{\mathcal{D},n}(\eta;\bm{\lambda})$ may not be
orthogonal polynomials).

\noindent
{\bf Remark 2}\,
The multi-indexed polynomials $P_{\mathcal{D},n}(\eta)$ ($n\geq M+1$) are
determined by the $3+2M$ term recurrence relations \eqref{RRP} with $M+1$
``initial data'' \cite{rrmiop},
\begin{equation}
  P_{\mathcal{D},0}(\eta),\,P_{\mathcal{D},1}(\eta),
  \,\ldots\,,\,P_{\mathcal{D},M}(\eta).
  \label{initdata}
\end{equation}
After calculating the initial data \eqref{initdata} by \eqref{MP:cPDndef} and
\eqref{cPDndef}, we can obtain $P_{\mathcal{D},n}(\eta)$ through
the $3+2M$ term recurrence relations \eqref{RRP}.
The calculation cost of this method is much less than the original determinant
expression \eqref{MP:cPDndef} and \eqref{cPDndef} for large $M$.

\subsection{Recurrence relations with constant coefficients}
\label{sec:rr_const}

We present $1+2L$ term recurrence relations with constant coefficients.
For the case-(1) multi-indexed Wilson and Askey-Wilson polynomials, such
recurrence relations are presented in \cite{rrmiop2} and shown in Appendix B
of \cite{rrmiop5}.
Since the derivation can be applied in the present case without difficulty,
we present only the results here.

We want to find the following recurrence relations,
\begin{equation*}
  X(\eta)P_{\mathcal{D},n}(\eta)
  =\sum_{k=-n}^Lr_{n,k}^{X,\mathcal{D}}P_{\mathcal{D},n+k}(\eta),
\end{equation*}
where $r_{n,k}^{X,\mathcal{D}}\,$'s are constants and $X(\eta)$ is some
polynomial of degree $L$ in $\eta$.
The overall normalization and the constant term of $X(\eta)$ are not important,
because the change of the former induces that of the overall normalization
of $r_{n,k}^{X,\mathcal{D}}$ and the shift of the latter induces that of
$r_{n,0}^{X,\mathcal{D}}$.

The sinusoidal coordinate $\eta(x)$ ($\eta(x)=x$ in the present case)
satisfies \cite{os14,rrmiop2}
\begin{equation}
  \frac{\eta(x-i\frac{\gamma}{2})^{n+1}-\eta(x+i\frac{\gamma}{2})^{n+1}}
  {\eta(x-i\frac{\gamma}{2})-\eta(x+i\frac{\gamma}{2})}
  =\sum_{k=0}^ng_n^{\prime\,(k)}\eta(x)^{n-k}
  \ \ (n\in\mathbb{Z}_{\geq 0}),
\end{equation}
where $g_n^{\prime\,(k)}$ is given by \cite{os32}
\begin{equation}
  g_n^{\prime\,(k)}=\theta(k:\text{even})\,
  (-1)^{\frac{k}{2}}2^{-k}\genfrac{(}{)}{0pt}{}{n+1}{k+1}.
\end{equation}
Here $\theta(P)$ is a step function for a proposition $P$ ; $\theta(P)=1$
for $P$\,:\,true, $\theta(P)=0$ for $P$\,:\,false.
For a polynomial $p(\eta)$ in $\eta$, let us define a polynomial in $\eta$,
$I[p](\eta)$, as follows:
\begin{equation}
  p(\eta)=\sum_{k=0}^na_k\eta^k\mapsto
  I[p](\eta)=\sum_{k=0}^{n+1}b_k\eta^k,
  \label{mapI}
\end{equation}
where $b_k$'s are defined by
\begin{equation}
  b_{k+1}=\frac{1}{g_k^{\prime\,(0)}}
  \Bigl(a_k-\sum_{j=k+1}^ng_j^{\prime\,(j-k)}b_{j+1}\Bigr)
  \ \ (k=n,n-1,\ldots,1,0),\quad
  b_0=0.
\end{equation}
The constant term of $I[p](\eta)$ is chosen to be zero.
It is easy to show that this polynomial $I[p](\eta)=P(\eta)$ satisfies
\begin{equation}
  \frac{\check{P}(x-i\frac{\gamma}{2})-\check{P}(x+i\frac{\gamma}{2})}
  {\eta(x-i\frac{\gamma}{2})-\eta(x+i\frac{\gamma}{2})}
  =\check{p}(x),
\end{equation}
where $\check{P}(x)=P(\eta(x))$ and $\check{p}(x)=p(\eta(x))$.

For the denominator polynomial
$\Xi_{\mathcal{D}}(\eta)$ and a polynomial in $\eta$, $Y(\eta)(\neq 0)$,
we set $X(\eta)=X^{\mathcal{D},Y}(\eta)$ as
\begin{equation}
  X(\eta)=I\bigl[\Xi_{\mathcal{D}}Y\bigr](\eta),\quad
  \deg X(\eta)=L=\ell_{\mathcal{D}}+\deg Y(\eta)+1,
  \label{X=I[XiY]}
\end{equation}
where $\Xi_{\mathcal{D}}Y$ means a polynomial
$(\Xi_{\mathcal{D}}Y)(\eta)=\Xi_{\mathcal{D}}(\eta)Y(\eta)$.
Note that $L\geq M+1$ because of $\ell_{\mathcal{D}}\geq M$.
The minimal degree one, which corresponds to $Y(\eta)=1$, is
\begin{equation}
  X_{\text{min}}(\eta)=I\bigl[\Xi_{\mathcal{D}}\bigr](\eta),\quad
  \deg X_{\text{min}}(\eta)=\ell_{\mathcal{D}}+1.
  \label{Xmin}
\end{equation}
Then we have the following theorem.
\begin{thm}
\label{thm:rr_const}
For any polynomial $Y(\eta)(\neq 0)$, we take
$X(\eta)=X^{\mathcal{D},Y}(\eta)$ as \eqref{X=I[XiY]}.
Then the multi-indexed Meixner-Pollaczek and continuous Hahn orthogonal
polynomials $P_{\mathcal{D},n}(\eta)$ satisfy $1+2L$ term recurrence
relations with constant coefficients:
\begin{equation}
  X(\eta)P_{\mathcal{D},n}(\eta)
  =\sum_{k=-L}^Lr_{n,k}^{X,\mathcal{D}}P_{\mathcal{D},n+k}(\eta),
  \label{XPthm}
\end{equation}
which hold for $n\in\mathbb{Z}_{\geq 0}$.
Here $r_{n,k}^{X,\mathcal{D}}$'s are constants.
\end{thm}
{\bf Remark 1}\,
By defining $r_{n,k}^{X,\mathcal{D}}=0$ for $n<0$, \eqref{XPthm} holds for
$n\in\mathbb{Z}$.

\noindent
{\bf Remark 2}\,
Any polynomial $X(\eta)$ giving the recurrence relations with constant
coefficients must have the form \eqref{X=I[XiY]} \cite{rrmiop2}.

\noindent
{\bf Remark 3}\,
Many parts of the derivation in \cite{rrmiop2} and Appendix B of \cite{rrmiop5}
are done algebraically,
but some parts use the orthogonality. So we can not conclude that Theorem
\ref{thm:rr_const} holds for $P_{\mathcal{D},n}(\eta;\bm{\lambda})$ with
any $\bm{\lambda}$ and $\mathcal{D}$.
However, explicit calculation for small $M$, $d_j$, $n$ and $\deg Y$ suggests
the following conjecture.
\begin{conj}
\label{conj:rr_const}
Theorem \ref{thm:rr_const} holds for $P_{\mathcal{D},n}(\eta;\bm{\lambda})$
with any $\bm{\lambda}$ and $\mathcal{D}$ (namely
$P_{\mathcal{D},n}(\eta;\bm{\lambda})$ may not be orthogonal polynomials).
\end{conj}

\noindent
{\bf Remark 4}\,
Direct verification of this theorem is rather straightforward for lower
$M$ and smaller $d_j$, $n$ and $\deg Y$, by a computer algebra system,
e.g.\! Mathematica.
The coefficients $r_{n,k}^{X,\mathcal{D}}$ are explicitly obtained for
small $d_j$ and $n$. However, to obtain the closed expression of
$r_{n,k}^{X,\mathcal{D}}$ for general $n$ is not an easy task even for
small $d_j$, and it is a different kind of problem.
We present some examples in \S\,\ref{sec:ex} and Appendix \ref{app:ex}.

\noindent
{\bf Remark 5}\,
Since $Y(\eta)$ is arbitrary, we obtain infinitely many recurrence relations.
However not all of them are independent. The relations among them are unclear.
For `$M=0$ case' (namely, ordinary orthogonal polynomials), it is trivial that
recurrence relations obtained from arbitrary $Y(\eta)$ ($\deg Y\geq 1$)
are derived by the three term recurrence relations.

\medskip

Let us consider some properties of the coefficient $r_{n,k}^{X,\mathcal{D}}$.
By using the orthogonality relation \eqref{orthocPDn} and the recurrence
relations \eqref{XPthm}, we obtain the relations among them,
\begin{equation}
  r_{n+k,-k}^{X,\mathcal{D}}
  =\frac{h_{\mathcal{D},n+k}}{h_{\mathcal{D},n}}\,r_{n,k}^{X,\mathcal{D}}
  \ \ (1\leq k\leq L).
  \label{prop_rnk}
\end{equation}
Explicit forms of $\frac{h_{\mathcal{D},n+k}}{h_{\mathcal{D},n}}$ with
$k\geq 0$ are
\begin{align}
  \text{MP :}&
  \ \ \frac{h_{\mathcal{D},n+k}}{h_{\mathcal{D},n}}
  =\frac{(n+2a)_k}{(n+1)_k}
  \prod_{j=1}^M\frac{n+2a-d_j-1+k}{n+2a-d_j-1},
  \label{MP:hD/hD}\\
  \text{cH :}&
  \ \ \frac{h_{\mathcal{D},n+k}}{h_{\mathcal{D},n}}
  =\frac{(n+a_1+a_1^*,n+a_1+a_2^*,n+a_2+a_1^*,n+a_2+a_2^*)_k}{(n+1,n+b_1-1)_k}
  \frac{2n+b_1-1}{2n+b_1-1+2k}\n
  &\phantom{\ \ \frac{h_{\mathcal{D},n+k}}{h_{\mathcal{D},n}}=}
  \times\prod_{j=1}^{M_{\I}}
  \frac{(n+a_1+a_1^*-1-d^{\I}_j+k)(n+a_2+a_2^*+d^{\I}_j+k)}
  {(n+a_1+a_1^*-1-d^{\I}_j)(n+a_2+a_2^*+d^{\I}_j)}\n
  &\phantom{\ \ \frac{h_{\mathcal{D},n+k}}{h_{\mathcal{D},n}}=}
  \times\prod_{j=1}^{M_{\II}}
  \frac{(n+a_2+a_2^*-1-d^{\II}_j+k)(n+a_1+a_1^*+d^{\II}_j+k)}
  {(n+a_2+a_2^*-1-d^{\II}_j)(n+a_1+a_1^*+d^{\II}_j)}.
  \label{hD/hD}
\end{align}
The values of $P_{\mathcal{D},n}(\eta)$ at some specific values $\eta_0$
are known explicitly as \eqref{MP:PDn(eta0)} and \eqref{PDn(eta0)}.
By substituting $\eta_0$ for $\eta$ in \eqref{XPthm}, we have
$X(\eta_0)P_{\mathcal{D},n}(\eta_0)
=\sum\limits_{k=-L}^Lr_{n,k}^{X,\mathcal{D}}P_{\mathcal{D},n+k}(\eta_0)$,
which gives
\begin{equation}
  r_{n,0}^{X,\mathcal{D}}=X(\eta_0)
  -\sum_{\genfrac{}{}{0pt}{1}{k=-L}{k\neq 0}}^L
  \frac{P_{\mathcal{D},n+k}(\eta_0)}{P_{\mathcal{D},n}(\eta_0)}
  r_{n,k}^{X,\mathcal{D}}.
  \label{rn0}
\end{equation}
Therefore it is sufficient to find $r_{n,k}^{X,\mathcal{D}}$ ($1\leq k\leq L$).
The top coefficient $r_{n,L}^{X,\mathcal{D}}$ is easily obtained by comparing
the highest degree terms,
\begin{equation}
  r_{n,L}^{X,\mathcal{D}}
  =\frac{c^Xc^P_{\mathcal{D},n}}{c^P_{\mathcal{D},n+L}},
\end{equation}
where $c^X$ is the coefficient of the highest term of
$X(\eta)=c^X\eta^L+(\text{lower order terms})$ and
$c^P_{\mathcal{D},n}$ is given by \eqref{MP:cPDn} and \eqref{cPDn}.
For later use, we provide a conjecture about $r_{n,0}^{X,\mathcal{D}}$.
\begin{conj}
\label{conj_rnk}
As a function of $n$, the coefficients $r_{n,0}^{X,\mathcal{D}}$ has the
following form,
\begin{equation}
  r_{n,0}^{X,\mathcal{D}}
  =-\frac{I(z)}{\prod_{j=1}^L\alpha_j(z)\alpha_{2L+1-j}(z)}\,
  \biggl|_{z=\mathcal{E}_n},\quad
  I(z):\text{a polynomial in $z$},
  \label{Iz}
\end{equation}
where $\alpha_j(z)\alpha_{2L+1-j}(z)$ will be given in \eqref{al*al}.
The degree of $I(z)$ is $\deg I=L$ for MP, $\deg I\leq 2L$ for cH.
\end{conj}
Note that this polynomial $I(z)$ is nothing to do with the map $I[\cdot]$ in
\eqref{mapI}.

\subsection{Examples}
\label{sec:ex}

For illustration, we present some examples of the coefficients
$r_{n,k}^{X,\mathcal{D}}$ of the recurrence relations \eqref{XPthm}
for multi-indexed orthogonal polynomials.
See Appendix \ref{app:ex} for non-orthogonal case.
Except for Ex.1 in \S\,\ref{sec:ex_MP},
we present only $r_{n,k}^{X,\mathcal{D}}$ ($1\leq k\leq L$), because
$r_{n,k}^{X,\mathcal{D}}$ ($-L\leq k\leq 0$) are obtained by
\eqref{prop_rnk}--\eqref{rn0}.

\subsubsection{multi-indexed Meixner-Pollaczek polynomials}
\label{sec:ex_MP}

\noindent
\underline{Ex.1} $\mathcal{D}=\{2\}$, $Y(\eta)=1$
($\Rightarrow\ell_{\mathcal{D}}=2,\,X(\eta)=X_{\text{min}}(\eta),\,L=3$):
7-term recurrence relations
\begin{align*}
  X(\eta)&=\frac{\eta}{12}\bigl(8\sin^2\phi\cdot\eta^2
  -6(2a-3)\sin 2\phi\cdot\eta
  +12a^2-24a+13+(12a^2-36a+23)\cos 2\phi\bigr),\\
  r_{n,3}^{X,\mathcal{D}}&=
  \frac{(n+1)_3}{12\sin\phi}\frac{2a+n-3}{2a+n},\quad
  r_{n,-3}^{X,\mathcal{D}}=
  \frac{(2a+n-3)_3}{12\sin\phi},\\
  r_{n,2}^{X,\mathcal{D}}&=
  -\frac12(n+1)_2\,(2a+n-3)\cot\phi,\quad
  r_{n,-2}^{X,\mathcal{D}}=
  -\frac12(2a+n-3)_3\,\cot\phi,\\
  r_{n,1}^{X,\mathcal{D}}&=
  \frac{(n+1)(2a+n-3)}{4\sin\phi}\bigl(4a+3n+2(2a+n-1)\cos 2\phi\bigr),\\
  r_{n,-1}^{X,\mathcal{D}}&=
  \frac{(2a+n-3)(2a+n-1)}{4\sin\phi}
  \bigl(4a+3n-3+2(2a+n-2)\cos 2\phi\bigr),\\
  r_{n,0}^{X,\mathcal{D}}&=
  -\frac{1}{6}\bigl((2a-3)(2a-1)(7a-1)+2(36a^2-60a+19)n+6(8a-7)n^2
  +10n^3\bigr)\cot\phi\\
  &\qquad
  +\frac{1}{24}(2a+2n-1)\bigl(4a(7a+5n-16)+33-22n+4n^2\bigr)\sin 2\phi.
\end{align*}
The polynomial $I(z)$ \eqref{Iz} is
\begin{align*}
  I(z)&=-48\sin\phi\sin 2\phi\Bigl(
  (4+\cos 2\phi)z^3
  +6\bigl(6a-5+2(a-1)\cos 2\phi\bigr)\sin\phi\cdot z^2\\
  &\quad
  +4\bigl(4(6a^2-9a+2)+(12a^2-24a+11)\cos 2\phi\bigr)\sin^2\phi\cdot z\\
  &\quad
  +(2a-3)(2a-1)\bigl(14a+7+(14a-11)\cos 2\phi\bigr)\sin^3\phi\Bigr).
\end{align*}

\noindent
\underline{Ex.2} $\mathcal{D}=\{1,2\}$, $Y(\eta)=1$
($\Rightarrow\ell_{\mathcal{D}}=2,\,X(\eta)=X_{\text{min}}(\eta),\,L=3$):
7-term recurrence relations
\begin{align*}
  X(\eta)&=\frac{2\sin\phi}{3}\eta\bigl(2\sin^2\phi\cdot\eta^2
  -3(a-1)\sin 2\phi\cdot\eta
  +3a^2-9a+7+(3a^2-6a+2)\cos 2\phi\bigr),\\
  r_{n,3}^{X,\mathcal{D}}&=
  \frac16(n+1)_3\frac{(2a+n-3)_2}{(2a+n)_2},\quad
  r_{n,2}^{X,\mathcal{D}}=
  -(n+1)_2\frac{(2a+n-3)_2}{2a+n}\cos\phi,\\
  r_{n,1}^{X,\mathcal{D}}&=
  \frac12(n+1)(2a+n-3)\bigl(4a+3n-4+2(2a+n-2)\cos 2\phi\bigr).
\end{align*}
The polynomial $I(z)$ \eqref{Iz} is 
\begin{align*}
  I(z)&=-96\sin^2\phi\sin 2\phi\Bigl(
  (4+\cos 2\phi)z^3
  +12(a-1)(3+\cos 2\phi)\sin\phi\cdot z^2\\
  &\quad
  +4\bigl(4(6a^2-12a+5)+(12a^2-24a+11)\cos 2\phi\bigr)\sin^2\phi\cdot z\\
  &\quad
  +8(a-1)\bigl(a(7a-11)+(7a^2-14a+6)\cos 2\phi\bigr)\sin^3\phi\Bigr).
\end{align*}

\noindent
\underline{Ex.3} $\mathcal{D}=\{2\}$, $Y(\eta)=\eta$
($\Rightarrow\ell_{\mathcal{D}}=2,\,L=4$):
9-term recurrence relations
\begin{align*}
  X(\eta)&=\frac{\eta}{24}\Bigl(
  12\sin^2\phi\cdot\eta^3
  -8(2a-3)\sin 2\phi\cdot\eta^2\\
  &\qquad
  +3\bigl(4a^2-8a+5+(4a^2-12a+7)\cos 2\phi\bigr)\eta
  -2(2a-3)\sin 2\phi\Bigr),\\
  r_{n,4}^{X,\mathcal{D}}&=
  \frac{(n+1)_4}{32\sin^2\phi}\frac{2a+n-3}{2a+n+1},\quad
  r_{n,3}^{X,\mathcal{D}}=
  -\frac{(n+1)_3\cos\phi}{12\sin^2\phi}\frac{2a+n-3}{2a+n}(5a+3n),\\
  r_{n,2}^{X,\mathcal{D}}&=
  \frac{(n+1)_2}{8\sin^2\phi}(2a+n-3)\bigl(2(2a+2n+1)
  +(4a+3n)\cos 2\phi\bigr),\\
  r_{n,1}^{X,\mathcal{D}}&=
  -\frac{(n+1)\cos\phi}{4\sin^2\phi}(2a+n-3)
  \bigl(4a(a+1)+(11a+1)n+5n^2\\
  &\qquad
  +2(a+n)(2a+n-1)\cos 2\phi\bigr).
\end{align*}
The polynomial $I(z)$ \eqref{Iz} is
\begin{align*}
  I(z)&=-192\sin^2\phi\Bigl(
  3(18+16\cos 2\phi+\cos 4\phi)z^4\\
  &\quad
  +4\bigl(36(4a-3)+4(34a-27)\cos 2\phi+(10a-9)\cos 4\phi\bigr)
  \sin\phi\cdot z^3\\
  &\quad
  +12\bigl(2(86a^2-119a+30)+8(22a^2-33a+10)\cos 2\phi
  +(16a^2-28a+11)\cos 4\phi\bigr)\sin^2\phi\cdot z^2\\
  &\quad
  +16\bigl(3(56a^3-98a^2+39a-9)+4(48a^3-96a^2+50a-9)\cos 2\phi\\
  &\qquad
  +(24a^3-60a^2+44a-9)\cos 4\phi\bigr)\sin^3\phi\cdot z\\
  &\quad
  +(2a-3)\bigl(3(2a+1)(68a^2-4a+1)+4(2a-1)(2a+1)(34a-3)\cos 2\phi\\
  &\qquad
  +(2a-1)(68a^2-80a+15)\cos 4\phi\bigr)\sin^4\phi\Bigr).
\end{align*}

\noindent
\underline{Ex.4} $\mathcal{D}=\{4\}$, $Y(\eta)=1$
($\Rightarrow\ell_{\mathcal{D}}=4,\,X(\eta)=X_{\text{min}}(\eta),\,L=5$):
11-term recurrence relations
\begin{align*}
  X(\eta)&=\frac{\eta}{960}\Bigl(
  128\sin^4\phi\cdot\eta^4
  -320(2a-5)\cos\phi\sin^3\phi\cdot\eta^3\\
  &\qquad
  +160\bigl(4a^2-16a+17+(4a^2-20a+23)\cos 2\phi\bigr)\sin^2\phi\cdot\eta^2\\
  &\qquad
  -80(2a-5)\bigl(4a^2-8a+5+(4a^2-20a+19)\cos 2\phi\bigr)
  \cos\phi\sin\phi\cdot\eta\\
  &\qquad
  +240a^4-1440a^3+3160a^2-3000a+1067\\
  &\qquad
  +4(80a^4-560a^3+1360a^2-1380a+511)\cos 2\phi\\
  &\qquad
  +(80a^4-800a^3+2760a^2-3800a+1689)\cos 4\phi\Bigr),\\
  r_{n,5}^{X,\mathcal{D}}&=
  \frac{(n+1)_5}{240\sin\phi}\frac{2a+n-5}{2a+n},\quad
  r_{n,4}^{X,\mathcal{D}}=
  -\frac{1}{24}(n+1)_4\,(2a+n-5)\cot\phi,\\
  r_{n,3}^{X,\mathcal{D}}&=
  \frac{(n+1)_3}{48\sin\phi}(2a+n-5)\bigl(8a+5n+4(2a+n-1)\cos 2\phi\bigr),\\
  r_{n,2}^{X,\mathcal{D}}&=
  -\frac16(n+1)_2\,(2a+n-5)(2a+n-1)\cot\phi
  \bigl(2a+2n+1+(2a+n-2)\cos 2\phi\bigr),\\
  r_{n,1}^{X,\mathcal{D}}&=
  \frac{n+1}{24\sin\phi}(2a+n-5)(2a+n-1)
  \bigl(12a^2+6a(4n-1)+10n(n-1)\\
  &\qquad
  +2(2a+n-2)(4a+5n)\cos 2\phi+(2a+n-3)_2\,\cos 4\phi\bigr).
\end{align*}
The polynomial $I(z)$ \eqref{Iz} is
\begin{align*}
  I(z)&=-3840\sin^3\phi\sin 2\phi\Bigl(
  (38+24\cos 2\phi+\cos 4\phi)z^5\\
  &\quad
  +10\bigl(8(7a-8)+4(10a-13)\cos 2\phi+(2a-3)\cos 4\phi\bigr)
  \sin\phi\cdot z^4\\
  &\quad
  +20\bigl(2(78a^2-173a+78)+4(32a^2-82a+47)\cos 2\phi\\
  &\qquad
  +(8a^2-24a+17)\cos 4\phi\bigr)\sin^2\phi\cdot z^3\\
  &\quad
  +40\bigl(4(2a-1)(25a^2-67a+35)+4(48a^3-180a^2+200a-65)\cos 2\phi\\
  &\qquad
  +(2a-3)(8a^2-24a+15)\cos 4\phi\bigr)\sin^3\phi\cdot z^2\\
  &\quad
  +32\bigl(280a^4-1100a^3+1310a^2-645a+131\\
  &\qquad
  +2(160a^4-760a^3+1180a^2-690a+119)\cos 2\phi\\
  &\qquad
  +(40a^4-240a^3+510a^2-450a+137)\cos 4\phi\bigr)\sin^4\phi\cdot z\\
  &\quad
  +(2a-5)(2a-1)\bigr(744a^3-828a^2+406a-297\\
  &\qquad
  +4(2a-3)(2a+1)(62a-57)\cos 2\phi\\
  &\qquad
  +(2a-3)(124a^2-352a+193)\cos 4\phi\bigr)\sin^5\phi\Bigr).
\end{align*}

We have also obtained 9-term recurrence relations for
$\mathcal{D}=\{1,2\}$ with non-minimal $X(\eta)$ ($Y(\eta)=\eta$), and
11-term recurrence relations for
$\mathcal{D}=\{1,4\},\{2,3\},\{1,2,4\},\{1,2,3,4\}$
with $X(\eta)=X_{\text{min}}(\eta)$
and $\mathcal{D}=\{2\},\{1,2\}$ with non-minimal $X(\eta)$ ($Y(\eta)=\eta^2$).
Since the explicit forms of $r_{n,k}^{X,\mathcal{D}}$ are somewhat lengthy,
we do not write down them here.

\subsubsection{multi-indexed continuous Hahn polynomials}
\label{sec:ex_cH}

We set $\sigma_1=a_1+a_1^*$, $\sigma_2=a_1a_1^*$,
$\sigma'_1=a_2+a_2^*$ and $\sigma'_2=a_2a_2^*$.

\noindent
\underline{Ex.1} $\mathcal{D}=\{2^{\I}\}$ ($M_{\I}=1,\,M_{\II}=0$), $Y(\eta)=1$
($\Rightarrow\ell_{\mathcal{D}}=2,\,X(\eta)=X_{\text{min}}(\eta),\,L=3$):
7-term recurrence relations
\begin{align}
  X(\eta)&=\frac{\eta}{24}\Bigl(4(\sigma_1-\sigma'_1-4)_2\,\eta^2
  +6i(\sigma_1-\sigma'_1-3)
  \bigl(a_1-a_1^*+2(a_1a_2^*-a_1^*a_2)+3(a_2-a_2^*)\bigr)\eta\n
  &\qquad
  +12+\bigl(36a_2(a_2+1)-2\sigma'_1-24\sigma'_2-7\bigr)\sigma_1
  +\bigl(1-12a_2(a_2+1)\bigr)\sigma_1^2\n
  &\qquad
  +24a_2(a_2+1)(a_1^2-3a_1+\sigma_2)-\sigma'_1(23\sigma'_1+17)
  -12a_1(a_1-3)\sigma'_1(\sigma'_1+1)\n
  &\qquad
  +24(a_1^2-3a_1+3+\sigma_2)\sigma'_2
  \Bigr),\n
  r_{n,3}^{X,\mathcal{D}}&=
  (\sigma_1-\sigma'_1-4)_2\,(n+1)_3
  \frac{(n+\sigma_1-3)(n+b_1-1)_3}{6(n+\sigma_1)(2n+b_1-1)_6},\n
  r_{n,2}^{X,\mathcal{D}}&=
  i(a_1-a_1^*-a_2+a_2^*)(\sigma_1-\sigma'_1-3)(b_1-2)(n+1)_2\n
  &\quad\times
  \frac{(n+\sigma_1-3)(n+\sigma'_1+2)(n+b_1-1)_2}
  {2(2n+b_1-2)_5\,(2n+b_1+4)},\n
  r_{n,1}^{X,\mathcal{D}}&=
  \frac{(n+\sigma_1-3)(n+\sigma'_1+2)(n+b_1-1)}{2(2n+b_1-3)_4\,(2n+b_1+2)_2}
  \times A,
  \label{A_Ex1_cH}
\end{align}
where $A$ is a polynomial of degree $5$ in $n$ and we present it in
Appendix\,\ref{app:AinEx1cH}, \eqref{AinEx1cH}, because it is a bit long.
The polynomial $I(z)$ of degree 4 \eqref{Iz} is omitted
because it has a lengthy expression.

The example with $\mathcal{D}=\{2^{\II}\}$ ($M_{\I}=0,\,M_{\II}=1$) and
$Y(\eta)=1$ can be obtained by exchanging $a_1$ and $a_2$.

We have also obtained 11-term recurrence relations for
$\mathcal{D}=\{4^{\I}\},\{4^{\II}\}$ with $X(\eta)=X_{\text{min}}(\eta)$.
Since the explicit forms of $r_{n,k}^{X,\mathcal{D}}$ are somewhat lengthy,
we do not write down them here.

\section{Generalized Closure Relations and Creation/An\-ni\-hi\-la\-tion
Operators}
\label{sec:gcr}

In this section we discuss the generalized closure relations and the
creation/annihilation operators of the multi-indexed Meixner-Pollaczek and
continuous Hahn idQM systems described by $\mathcal{H}_{\mathcal{D}}$
\eqref{HD}.

First let us recapitulate the essence of the (generalized) closure relation
\cite{rrmiop4}.
The closure relation of order $K$ is an algebraic relation between a
Hamiltonian $\mathcal{H}$ and some operator
$X$ ($=X(\eta(x))=\check{X}(x)$) \cite{rrmiop4}:
\begin{equation}
  (\text{ad}\,\mathcal{H})^KX
  =\sum_{i=0}^{K-1}(\text{ad}\,\mathcal{H})^iX\cdot R_i(\mathcal{H})
  +R_{-1}(\mathcal{H}),
  \label{crK}
\end{equation}
where $(\text{ad}\,\mathcal{H})X=[\mathcal{H},X]$,
$(\text{ad}\,\mathcal{H})^0X=X$ and $R_i(z)=R^X_i(z)$ is
a polynomial in $z$.
The original closure relation \cite{os7,os12} corresponds to $K=2$.
Since the closure relation of order $K$ implies that of order $K'>K$,
we are interested in the smallest integer $K$ satisfying \eqref{crK}.
We assume that the matrix $A=(a_{ij})_{1\leq i,j\leq K}$
($a_{i+1,i}=1$ ($1\leq i\leq K-1$), $a_{i+1,K}=R_i(z)$ ($0\leq i\leq K-1$),
$a_{ij}=0$ (others)) has $K$ distinct real non-vanishing eigenvalues
$\alpha_i=\alpha_i(z)$ for $z\geq 0$, which are indexed in decreasing order
$\alpha_1(z)>\alpha_2(z)>\cdots>\alpha_K(z)$.
Then we obtain the exact Heisenberg solution of $X$,
\begin{equation}
  X_{\text{H}}(t)\eqdef e^{i\mathcal{H}t}Xe^{-i\mathcal{H}t}
  =\sum_{n=0}^{\infty}\frac{(it)^n}{n!}(\text{ad}\,\mathcal{H})^nX
  =\sum_{j=1}^Ka^{(j)}e^{i\alpha_j(\mathcal{H})t}
  -R_{-1}(\mathcal{H})R_0(\mathcal{H})^{-1}.
  \label{X(t)}
\end{equation}
Here $a^{(j)}=a^{(j)}(\mathcal{H},X)$ ($1\leq j\leq K$) are creation or
annihilation operators,
\begin{equation}
  a^{(j)}=\Bigl(\sum_{i=1}^K(\text{ad}\,\mathcal{H})^{i-1}X\cdot
  p_{ij}(\mathcal{H})
  +R_{-1}(\mathcal{H})\alpha_j(\mathcal{H})^{-1}\Bigr)
  \prod_{\genfrac{}{}{0pt}{1}{k=1}{k\neq j}}^K
  (\alpha_j(\mathcal{H})-\alpha_k(\mathcal{H}))^{-1},
  \label{a(j)}
\end{equation}
where $p_{ij}(z)$ ($1\leq i,j\leq K$) are
\begin{equation}
  p_{ij}(z)=\alpha_j(z)^{K-i}-\sum_{k=1}^{K-i}R_{K-k}(z)\,\alpha_j(z)^{K-i-k}.
\end{equation}

Let us consider the idQM systems described by the multi-indexed
Meixner-Pollaczek and continuous Hahn polynomials.
The Hamiltonian is $\mathcal{H}_{\mathcal{D}}$ $\eqref{HD}$ and a candidate
of the operator $X$ is a polynomial $X(\eta(x))=\check{X}(x)$ discussed in
\S\,\ref{sec:rr_const}.
The closure relation \eqref{crK} is now
\begin{equation}
  (\text{ad}\,\mathcal{H}_{\mathcal{D}})^KX
  =\sum_{i=0}^{K-1}(\text{ad}\,\mathcal{H}_{\mathcal{D}})^i
  X\cdot R_i(\mathcal{H}_{\mathcal{D}})
  +R_{-1}(\mathcal{H}_{\mathcal{D}}).
  \label{crHD}
\end{equation}
{}From the form of $\mathcal{H}_{\mathcal{D}}$,
the polynomials $R_i(z)=R^X_i(z)$ have at most the following degrees,
\begin{equation}
  R_i(z)=\sum_{j=0}^{K-i}r_i^{(j)}z^j
  \ \ (0\leq i\leq K-1),\quad
  R_{-1}(z)=\sum_{j=0}^Kr_{-1}^{(j)}z^j,
\end{equation}
where $r_i^{(j)}=r_i^{X(j)}$ are coefficients.

Let us define $\alpha_j(z)$ ($1\leq j\leq 2L$) as follows:
\begin{align}
  \text{MP}:\ \ &\alpha_j(z)=\left\{
  \begin{array}{ll}
  2(L+1-j)\sin\phi&(1\leq j\leq L)\\[4pt]
  -2(j-L)\sin\phi&(L+1\leq j\leq 2L)
  \end{array}\right.,
  \label{alphaj_R}\\
  \text{cH}:\ \ &\alpha_j(z)=\left\{
  \begin{array}{ll}
  (L+1-j)^2+(L+1-j)\sqrt{4z+(b_1-1)^2}&(1\leq j\leq L)\\[4pt]
  (j-L)^2-(j-L)\sqrt{4z+(b_1-1)^2}&(L+1\leq j\leq 2L)
  \end{array}\right..
  \label{alphaj_cH}
\end{align}
For MP, $\alpha_j(z)$'s are constant functions.
The pair of $\alpha_j(z)$ and $\alpha_{2L+1-j}(z)$ ($1\leq j\leq L$) satisfies
\begin{align}
  \alpha_j(z)+\alpha_{2L+1-j}(z)&=\left\{
  \begin{array}{ll}
  0&:\text{MP}\\[2pt]
  2(L+1-j)^2&:\text{cH}\\
  \end{array}\right.,
  \label{al+al}\\[2pt]
  \alpha_j(z)\alpha_{2L+1-j}(z)&=\left\{
  \begin{array}{ll}
  -4(L+1-j)^2\sin^2\phi&:\text{MP}\\[2pt]
  (L+1-j)^2\bigl((L+1-j)^2-4z-(b_1-1)^2\bigr)&:\text{cH}
  \end{array}\right..
  \label{al*al}
\end{align}
These $\alpha_j(z)$ satisfy
\begin{equation}
  \alpha_1(z)>\alpha_2(z)>\cdots>\alpha_L(z)>0
  >\alpha_{L+1}(z)>\alpha_{L+2}(z)>\cdots>\alpha_{2L}(z)\ \ (z\geq 0),
  \label{alphajorder}
\end{equation}
for $0<\phi<\pi$ (MP) and $b_1>2L$ (cH).
We remark that $\alpha_j(\mathcal{E}_n)$ for cH is square root free,
$\sqrt{4\mathcal{E}_n+(b_1-1)^2}=2n+b_1-1$.
It is easy to show the following:
\begin{equation}
  \alpha_{j}(\mathcal{E}_n)=\left\{
  \begin{array}{ll}
  \mathcal{E}_{n+L+1-j}-\mathcal{E}_n>0&(1\leq j\leq L)\\[2pt]
  \mathcal{E}_{n-(j-L)}-\mathcal{E}_n<0&(L+1\leq j\leq 2L)
  \end{array}\right..
  \label{alphajEn}
\end{equation}

Like the Wilson and Askey-Wilson cases \cite{rrmiop4}, we conjecture the
following.
\begin{conj}
\label{conj_cr}
Take $X(\eta)$ as Theorem \ref{thm:rr_const} and take
$R_i(z)$ $(-1\leq i\leq 2L-1)$ as follows:
\begin{align}
  R_i(z)&=(-1)^{i+1}\!\!\!\!\!\!\!\!\!\!\!
  \sum_{1\leq j_1<j_2<\cdots<j_{2L-i}\leq 2L}\!\!\!\!\!\!\!\!\!\!\!\!\!\!
  \alpha_{j_1}(z)\alpha_{j_2}(z)\cdots\alpha_{j_{2L-i}}(z)
  \ \ (0\leq i\leq 2L-1),
  \label{Riz}\\
  R_{-1}(z)&=-I(z),
\end{align}
where $I(z)$ is given by \eqref{Iz}.
Then the closure relation of order $K=2L$ \eqref{crHD} holds.
\end{conj}
We remark that $R_i(z)$ in \eqref{Riz} are indeed polynomials in $z$,
because RHS of \eqref{Riz} are symmetric under the exchange of $\alpha_j$
and $\alpha_{2L+1-j}$ and their sum and product are polynomials in $z$,
\eqref{al+al}--\eqref{al*al}.
For MP, $R_i(z)$ ($0\leq i\leq 2L-1$) are constant functions;
$R_{2j}(z)=(-1)^{L-j-1}(2\sin\phi)^{2L-2j}\times(\text{positive integer})$
and $R_{2j+1}(z)=0$ ($0\leq j\leq L-1$).
Since $R_i(z)$ ($0\leq i\leq 2L-1$) are expressed in terms of $\alpha_j(z)$,
they do not depend on $\mathcal{D}$ and $X$ (except for $\deg X=L$).
Only $R_{-1}(z)$ depends on $\mathcal{D}$ and $X$.
For `$L=1$ case', namely the original system ($\mathcal{D}=\emptyset$,
$\ell_{\mathcal{D}}=0$, $\Xi_{\mathcal{D}}(\eta)=1$,
$X(\eta)=X_{\text{min}}(\eta)=\eta$), this generalized closure relation
reduces to the original closure relation \cite{os7}.
Direct verification of this conjecture is straightforward for lower
$M$ and smaller $d_j$ and $\deg Y$, by a computer algebra system.

If Conjecture \ref{conj_cr} is true,
we have the exact Heisenberg operator solution $X_{\text{H}}(t)$ \eqref{X(t)}
and the creation/annihilation operators $a^{(j)}=a^{\mathcal{D},X(j)}$
\eqref{a(j)}.
Action of \eqref{X(t)} on $\phi_{\mathcal{D}\,n}(x)$ \eqref{phiDn} is
\begin{equation*}
  e^{i\mathcal{H}_\mathcal{D}t}X
  e^{-i\mathcal{H}_\mathcal{D}t}\phi_{\mathcal{D}\,n}(x)
  =\sum_{j=1}^{2L}e^{i\alpha_j(\mathcal{E}_n)t}a^{(j)}\phi_{\mathcal{D}\,n}(x)
  -R_{-1}(\mathcal{E}_n)R_0(\mathcal{E}_n)^{-1}\phi_{\mathcal{D}\,n}(x).
\end{equation*}
On the other hand the LHS turns out to be
\begin{align*}
  e^{i\mathcal{H}_\mathcal{D}t}X
  e^{-i\mathcal{H}_\mathcal{D}t}\phi_{\mathcal{D}\,n}(x)
  &=e^{i\mathcal{H}_\mathcal{D}t}X
  e^{-i\mathcal{E}_nt}\phi_{\mathcal{D}\,n}(x)
  =e^{-i\mathcal{E}_nt}e^{i\mathcal{H}_\mathcal{D}t}
  \sum_{k=-L}^Lr_{n,k}^{X,\mathcal{D}}\phi_{\mathcal{D}\,n+k}(x)\n
  &=\sum_{k=-L}^Le^{i(\mathcal{E}_{n+k}-\mathcal{E}_n)t}\,
  r_{n,k}^{X,\mathcal{D}}\phi_{\mathcal{D}\,n+k}(x),
\end{align*}
where we have used \eqref{XPthm}.
Comparing these $t$-dependence (we assume $b_1>2L$ for cH),
we obtain \eqref{alphajEn} and
\begin{align}
  &a^{(j)}\phi_{\mathcal{D}\,n}(x)=\left\{
  \begin{array}{ll}
  r_{n,L+1-j}^{X,\mathcal{D}}\phi_{\mathcal{D}\,n+L+1-j}(x)
  &(1\leq j\leq L)\\[6pt]
  r_{n,-(j-L)}^{X,\mathcal{D}}\phi_{\mathcal{D}\,n-(j-L)}(x)
  &(L+1\leq j\leq 2L)
  \end{array}\right.,
  \label{ajphiDn}\\[2pt]
  &-R_{-1}(\mathcal{E}_n)R_0(\mathcal{E}_n)^{-1}
  =r_{n,0}^{X,\mathcal{D}},
  \label{Rm1En}
\end{align}
where $r_{n,k}^{X,\mathcal{D}}=0$ for $n+k<0$.
Note that \eqref{Rm1En} is consistent with Conjecture \ref{conj_rnk}.
Therefore $a^{(j)}$ ($1\leq j\leq L$) and $a^{(j)}$ ($L+1\leq j\leq 2L$)
are creation and annihilation operators, respectively.
Among them, $a^{(L)}$ and $a^{(L+1)}$ are fundamental,
$a^{(L)}\phi_{\mathcal{D},n}(x)\propto\phi_{\mathcal{D}\,n+1}(x)$ and
$a^{(L+1)}\phi_{\mathcal{D},n}(x)\propto\phi_{\mathcal{D}\,n-1}(x)$.
Furthermore, $X=X_{\text{min}}$ case is the most basic.

By the similarity transformation (see \eqref{tHD}),
the closure relation \eqref{crHD} becomes
\begin{equation}
  (\text{ad}\,\widetilde{\mathcal{H}}_{\mathcal{D}})^KX
  =\sum_{i=0}^{K-1}(\text{ad}\,\widetilde{\mathcal{H}}_{\mathcal{D}})^i
  X\cdot R_i(\widetilde{\mathcal{H}}_{\mathcal{D}})
  +R_{-1}(\widetilde{\mathcal{H}}_{\mathcal{D}}),
  \label{crtHD}
\end{equation}
and the creation/annihilation operators for eigenpolynomials can be obtained,
\begin{align}
  &\tilde{a}^{(j)}\eqdef\psi_{\mathcal{D}}(x)^{-1}\circ
  a^{(j)}(\mathcal{H}_{\mathcal{D}},X)\circ\psi_{\mathcal{D}}(x)
  =a^{(j)}(\widetilde{\mathcal{H}}_{\mathcal{D}},X),
  \label{ajt}\\
  &\tilde{a}^{(j)}\check{P}_{\mathcal{D},n}(x)=\left\{
  \begin{array}{ll}
  r_{n,L+1-j}^{X,\mathcal{D}}\check{P}_{\mathcal{D},n+L+1-j}(x)
  &(1\leq j\leq L)\\[6pt]
  r_{n,-(j-L)}^{X,\mathcal{D}}\check{P}_{\mathcal{D},n-(j-L)}(x)
  &(L+1\leq j\leq 2L)
  \end{array}\right..
  \label{ajtPDn}
\end{align}

\noindent
{\bf Remark}\,
Since the closure relations \eqref{crHD} (or \eqref{crtHD}) is an algebraic
relation between $\mathcal{H}_{\mathcal{D}}$ (or
$\widetilde{\mathcal{H}}_{\mathcal{D}}$) and $X$, it is expected to hold
even when $\mathcal{H}_{\mathcal{D}}$ (or
$\widetilde{\mathcal{H}}_{\mathcal{D}}$) is singular.
Especially we conjecture that $\tilde{a}^{(j)}$ \eqref{ajt} are the
creation/annihilation operators for eigenpolynomials
$\check{P}_{\mathcal{D},n}(x)$ with any $\mathcal{D}$, \eqref{ajtPDn}, which
can be verified by direct calculation for small $d_j$, $n$ and $\deg\,Y$.

\medskip

For an illustration, we present an example.
Let us consider Ex.1 in \S\,\ref{sec:ex_MP}.
The denominator polynomial $\Xi_{\mathcal{D}}(\eta)$ is
\begin{equation*}
  \Xi_{\mathcal{D}}(\eta)=(1-\cos 2\phi)\eta^2
  -(2a-3)\sin 2\phi\cdot\eta+(a-1)\bigl(a-1+(a-2)\cos 2\phi\bigr),
\end{equation*}
and $R_i(z)$ are
\begin{align*}
  &R_5(z)=R_3(z)=R_1(z)=0,\n
  &R_4(z)=56\sin^2\phi,\quad R_2(z)=-784\sin^4\phi,\quad
  R_0(z)=2304\sin^6\phi,\\
  &R_{-1}(z)=48\sin\phi\,\sin 2\phi\Bigl(
  (4+\cos2\phi)z^3+6\bigl(6a-5+2(a-1)\cos2\phi\bigl)\sin\phi\cdot z^2\n
  &\phantom{R_{-1}(z)=}
  +4\bigl(4(6a^2-9a+2)+(12a^2-24a+11)\cos2\phi\bigr)\sin^2\phi\cdot z\n
  &\phantom{R_{-1}(z)=}
  +(2a-3)(2a-1)\bigl(14a+7+(14a-11)\cos2\phi\bigr)\sin^3\phi\Bigr).
\end{align*}
The creation/annihilation operators for eigenpolynomials $\tilde{a}^{(j)}$ are
\begin{align*}
  \tilde{a}^{(1)}&=\frac{1}{7680}\Bigl(384X
  +\frac{64}{\sin\phi}(\text{ad}\,\widetilde{\mathcal{H}})X
  -\frac{120}{\sin^2\phi}(\text{ad}\,\widetilde{\mathcal{H}})^2X
  -\frac{20}{\sin^3\phi}(\text{ad}\,\widetilde{\mathcal{H}})^3X\\
  &\qquad
  +\frac{6}{\sin^4\phi}(\text{ad}\,\widetilde{\mathcal{H}})^4X
  +\frac{1}{\sin^5\phi}(\text{ad}\,\widetilde{\mathcal{H}})^5X
  +\frac{1}{6\sin^6\phi}R_{-1}(\widetilde{\mathcal{H}})\Bigr),\\
  \tilde{a}^{(2)}&=\frac{1}{1920}\Bigl(-576X
  -\frac{144}{\sin\phi}(\text{ad}\,\widetilde{\mathcal{H}})X
  +\frac{160}{\sin^2\phi}(\text{ad}\,\widetilde{\mathcal{H}})^2X
  +\frac{40}{\sin^3\phi}(\text{ad}\,\widetilde{\mathcal{H}})^3X\\
  &\qquad
  -\frac{4}{\sin^4\phi}(\text{ad}\,\widetilde{\mathcal{H}})^4X
  -\frac{1}{\sin^5\phi}(\text{ad}\,\widetilde{\mathcal{H}})^5X
  -\frac{1}{4\sin^6\phi}R_{-1}(\widetilde{\mathcal{H}})\Bigr),\\
  \tilde{a}^{(3)}&=\frac{1}{1536}\Bigl(1152X
  +\frac{576}{\sin\phi}(\text{ad}\,\widetilde{\mathcal{H}})X
  -\frac{104}{\sin^2\phi}(\text{ad}\,\widetilde{\mathcal{H}})^2X
  -\frac{52}{\sin^3\phi}(\text{ad}\,\widetilde{\mathcal{H}})^3X\\
  &\qquad
  +\frac{2}{\sin^4\phi}(\text{ad}\,\widetilde{\mathcal{H}})^4X
  +\frac{1}{\sin^5\phi}(\text{ad}\,\widetilde{\mathcal{H}})^5X
  +\frac{1}{2\sin^6\phi}R_{-1}(\widetilde{\mathcal{H}})\Bigr),\\
  \tilde{a}^{(4)}&=\frac{1}{1536}\Bigl(1152X
  -\frac{576}{\sin\phi}(\text{ad}\,\widetilde{\mathcal{H}})X
  -\frac{104}{\sin^2\phi}(\text{ad}\,\widetilde{\mathcal{H}})^2X
  +\frac{52}{\sin^3\phi}(\text{ad}\,\widetilde{\mathcal{H}})^3X\\
  &\qquad
  +\frac{2}{\sin^4\phi}(\text{ad}\,\widetilde{\mathcal{H}})^4X
  -\frac{1}{\sin^5\phi}(\text{ad}\,\widetilde{\mathcal{H}})^5X
  +\frac{1}{2\sin^6\phi}R_{-1}(\widetilde{\mathcal{H}})\Bigr),\\
  \tilde{a}^{(5)}&=\frac{1}{1920}\Bigl(-576X
  +\frac{144}{\sin\phi}(\text{ad}\,\widetilde{\mathcal{H}})X
  +\frac{160}{\sin^2\phi}(\text{ad}\,\widetilde{\mathcal{H}})^2X
  -\frac{40}{\sin^3\phi}(\text{ad}\,\widetilde{\mathcal{H}})^3X\\
  &\qquad
  -\frac{4}{\sin^4\phi}(\text{ad}\,\widetilde{\mathcal{H}})^4X
  +\frac{1}{\sin^5\phi}(\text{ad}\,\widetilde{\mathcal{H}})^5X
  -\frac{1}{4\sin^6\phi}R_{-1}(\widetilde{\mathcal{H}})\Bigr),\\
  \tilde{a}^{(6)}&=\frac{1}{7680}\Bigl(384X
  -\frac{64}{\sin\phi}(\text{ad}\,\widetilde{\mathcal{H}})X
  -\frac{120}{\sin^2\phi}(\text{ad}\,\widetilde{\mathcal{H}})^2X
  +\frac{20}{\sin^3\phi}(\text{ad}\,\widetilde{\mathcal{H}})^3X\\
  &\qquad
  +\frac{6}{\sin^4\phi}(\text{ad}\,\widetilde{\mathcal{H}})^4X
  -\frac{1}{\sin^5\phi}(\text{ad}\,\widetilde{\mathcal{H}})^5X
  +\frac{1}{6\sin^6\phi}R_{-1}(\widetilde{\mathcal{H}})\Bigr).
\end{align*}
Here $(\text{ad}\,\widetilde{\mathcal{H}})^iX$ are
\begin{align*}
  (\text{ad}\,\widetilde{\mathcal{H}})X&=
  V'_{\mathcal{D}}(x)\bigl(X(x-i\gamma)-X(x)\bigr)e^{\gamma p}
  +V_{\mathcal{D}}^{\prime\,*}(x)\bigl(X(x+i\gamma)-X(x)\bigr)e^{-\gamma p},\\
  (\text{ad}\,\widetilde{\mathcal{H}})^2X&=
  V'_{\mathcal{D}}(x)V'_{\mathcal{D}}(x-i\gamma)
  \bigl(X(x)-2X(x-i\gamma)+X(x-2i\gamma)\bigr)e^{2\gamma p}\\
  &\quad
  +V'_{\mathcal{D}}(x)\bigl(V_{\mathcal{D}}(x)+V_{\mathcal{D}}^*(x)
  -V_{\mathcal{D}}(x-i\gamma)-V_{\mathcal{D}}^*(x-i\gamma)\bigr)
  \bigl(X(x)-X(x-i\gamma)\bigr)e^{\gamma p}\\
  &\quad
  +2V'_{\mathcal{D}}(x)V_{\mathcal{D}}^{\prime\,*}(x-i\gamma)
  \bigl(X(x)-X(x-i\gamma)\bigr)
  +2V_{\mathcal{D}}^{\prime\,*}(x)V'_{\mathcal{D}}(x+i\gamma)
  \bigl(X(x)-X(x+i\gamma)\bigr)\\
  &\quad
  +V_{\mathcal{D}}^{\prime\,*}(x)
  \bigl(V_{\mathcal{D}}(x)+V_{\mathcal{D}}^*(x)
  -V_{\mathcal{D}}(x+i\gamma)-V_{\mathcal{D}}^*(x+i\gamma)\bigr)
  \bigl(X(x)-X(x+i\gamma)\bigr)e^{-\gamma p}\\
  &\quad
  +V_{\mathcal{D}}^{\prime\,*}(x)V_{\mathcal{D}}^{\prime\,*}(x+i\gamma)
  \bigl(X(x)-2X(x+i\gamma)+X(x+2i\gamma)\bigr)e^{-2\gamma p},
\end{align*}
and so on (we omit them because they are somewhat lengthy).
By direct calculation, the relations \eqref{ajtPDn} can be verified for
small $n$.

\section{Summary and Comments}
\label{sec:summary}

Following the preceding papers on the case-(1) multi-indexed orthogonal
polynomials (Laguerre and Jacobi cases in oQM \cite{rrmiop,rrmiop2,rrmiop3},
Wilson and Askey-Wilson cases in idQM \cite{rrmiop,rrmiop2,rrmiop5} and
Racah and $q$-Racah cases in rdQM \cite{rrmiop5}), we have discussed
the recurrence relations for the case-(1) multi-indexed Meixner-Pollaczek
and continuous Hahn orthogonal polynomials in idQM, whose physical range of
the coordinate is the whole real line.
The $3+2M$ term recurrence relations with variable dependent coefficients
\eqref{RRP} (Theorem\,\ref{thm:rr_var}) provide an efficient method to
calculate the multi-indexed MP and cH polynomials.
The $1+2L$ term ($L\geq M+1$) recurrence relations with constant coefficients
\eqref{XPthm} (Theorem\,\ref{thm:rr_const}), and their examples are presented.
Since $Y(\eta)$ is arbitrary, we obtain infinitely many recurrence relations.
Not all of them are independent, but the relations among them are unclear.
To clarify their relations is an important problem.
Corresponding to the recurrence relations with constant coefficients,
the idQM systems described by the multi-indexed MP and cH orthogonal
polynomials satisfy the generalized closure relations \eqref{crHD}
(Conjecture\,\ref{conj_cr}), from which the creation and
annihilation operators are obtained.
There are many creation and annihilation operators and it is an interesting
problem to study their relations.

The Hamiltonian of the deformed system is determined by the denominator
polynomial $\Xi_{\mathcal{D}}(\eta)$, whose degree is $\ell_{\mathcal{D}}$
\eqref{lD}.
There is no restriction on $\ell_{\mathcal{D}}$ for L, J, W and AW cases,
whereas the degree $\ell_{\mathcal{D}}$ must be even for MP and cH cases,
in order that the deformed Hamiltonian is hermitian.
This is because the physical range of the coordinate of the deformed MP and
cH systems is the whole real line, see \eqref{nozero}.
The range of the coordinate of the harmonic oscillator, whose eigenstates are
described by the Hermite polynomial, is also the whole real line, but
the case-(1) multi-indexed Hermite orthogonal polynomials do not exist.
In \S\,\ref{sec:miopdef} we have defined the multi-indexed MP and cH
polynomials for any index set $\mathcal{D}$, namely $\ell_{\mathcal{D}}$ may
be odd and they may not be orthogonal polynomials.
The recurrence relations with variable dependent coefficients \eqref{RRP}
for the multi-indexed MP and cH polynomials hold even for non-orthogonal case.
We conjecture that the recurrence relations with constant coefficients
\eqref{XPthm}, the generalized closure relations \eqref{crtHD} and the
creation/annihilation operators \eqref{ajtPDn} also hold even for
non-orthogonal case.

\section*{Acknowledgments}

This work was supported by JSPS KAKENHI Grant Numbers JP19K03667.

\bigskip
\appendix
\section{Discrete Quantum Mechanics With Pure Imaginary Shifts and
Deformed Systems}
\label{app:idQM}

In this Appendix we recapitulate the discrete quantum mechanics with pure
imaginary shifts (idQM) and deformed systems \cite{os13,os24,os27,idQMcH}.

The dynamical variables of idQM are the real coordinate $x$ ($x_1<x<x_2$)
and the conjugate momentum $p=-i\partial_x$, which are governed by the
following factorized positive semi-definite Hamiltonian:
\begin{align}
  &\mathcal{H}\eqdef\sqrt{V(x)}\,e^{\gamma p}\sqrt{V^*(x)}
  +\!\sqrt{V^*(x)}\,e^{-\gamma p}\sqrt{V(x)}
  -V(x)-V^*(x)=\mathcal{A}^{\dagger}\mathcal{A},
  \label{H}\\
  &\mathcal{A}\eqdef i\bigl(e^{\frac{\gamma}{2}p}\sqrt{V^*(x)}
  -e^{-\frac{\gamma}{2}p}\sqrt{V(x)}\,\bigr),\quad
  \mathcal{A}^{\dagger}\eqdef-i\bigl(\sqrt{V(x)}\,e^{\frac{\gamma}{2}p}
  -\sqrt{V^*(x)}\,e^{-\frac{\gamma}{2}p}\bigr).
\end{align}
Here the potential function $V(x)$ is an analytic function of $x$ and
$\gamma$ is a real constant.
The $*$-operation on an analytic function $f(x)=\sum_na_nx^n$
($a_n\in\mathbb{C}$) is defined by $f^*(x)=\sum_na_n^*x^n$, in which
$a_n^*$ is the complex conjugation of $a_n$.
Since the momentum operator appears in exponentiated forms,
the Schr\"odinger equation
\begin{equation}
  \mathcal{H}\phi_n(x)=\mathcal{E}_n\phi_n(x)
  \ \ (n=0,1,2,\ldots),
  \label{Hphin=}
\end{equation}
is an analytic difference equation with pure imaginary shifts instead
of a differential equation.
We consider those systems which have a
square-integrable groundstate together with an infinite number of
discrete energy levels:
$0=\mathcal{E}_0 <\mathcal{E}_1 < \mathcal{E}_2 < \cdots$.
The orthogonality relation reads
\begin{equation}
  (\phi_n,\phi_m)\eqdef
  \int_{x_1}^{x_2}\!\!dx\,\phi_n^*(x)\phi_m(x)=h_n\delta_{nm}
  \ \ (n,m=0,1,2,\ldots),\quad 0<h_n<\infty.
  \label{(phin,phim)}
\end{equation}
The eigenfunctions $\phi_n(x)$ can be chosen `real', $\phi_n^*(x)=\phi_n(x)$,
and the groundstate wavefunction $\phi_0(x)$ is determined as the zero
mode of the operator $\mathcal{A}$, $\mathcal{A}\phi_0(x)=0$.
The norm of a function $f(x)$ is $|\!|f|\!|\eqdef(f,f)^{\frac12}$.

The Hamiltonian $\mathcal{H}$ should be hermitian.
{}From its form $\mathcal{H}=\mathcal{A}^{\dagger}\mathcal{A}$, it is formally
hermitian, $\mathcal{H}^{\dagger}=(\mathcal{A}^{\dagger}\mathcal{A})^{\dagger}
=(\mathcal{A})^{\dagger}(\mathcal{A}^{\dagger})^{\dagger}
=\mathcal{A}^{\dagger}\mathcal{A}=\mathcal{H}$.
However, the true hermiticity is defined in terms of the inner product,
$(f_1,\mathcal{H}f_2)=(\mathcal{H}f_1,f_2)$ \cite{os13,os14,os27}.
To show the hermiticity of $\mathcal{H}$, singularities of some functions
in the rectangular domain $D_{\gamma}$ are important.
Here $D_{\gamma}$ is defined by \cite{os27}
\begin{equation}
  D_{\gamma}\eqdef\bigl\{x\in\mathbb{C}\bigm|x_1\leq\text{Re}\,x\leq x_2,
  |\text{Im}\,x|\leq\tfrac12|\gamma|\bigr\}.
  \label{Dgamma}
\end{equation}

The Meixner-Pollaczek (MP), continuous Hahn (cH), Wilson (W), Askey-Wilson
(AW) polynomials etc.\ are members of the Askey-scheme of the (basic)
hypergeometric orthogonal polynomials and satisfy the second order analytic
difference equation with pure imaginary shifts \cite{kls}.
These orthogonal polynomials can be studied in the framework of idQM,
in which they appear as part of the eigenfunction as follows:
\begin{equation}
  \phi_n(x)=\phi_0(x)\check{P}_n(x),\quad
  \check{P}_n(x)\eqdef P_n\bigl(\eta(x)\bigr)\quad
  (n=0,1,2,\ldots).
  \label{phin}
\end{equation}
Here $\eta(x)$ is a sinusoidal coordinate \cite{os7,os14} and
$P_n(\eta)$ is a orthogonal polynomial of degree $n$ in $\eta$.
The orthogonality relation \eqref{(phin,phim)} gives that of $\check{P}_n(x)$,
\begin{equation}
  \int_{x_1}^{x_2}\!\!dx\,\phi_0(x)^2
  \check{P}_n(x)\check{P}_m(x)
  =h_n\delta_{nm}\ \ (n,m=0,1,2,\ldots).
  \label{orthocPn}
\end{equation}
We call this idQM system by the name of the orthogonal polynomial:
MP system, cH system, W system, AW system etc.
These idQM systems have the property of shape invariance, which is a
sufficient condition for exact solvability.
Concrete idQM systems have a set of parameters
$\bm{\lambda}=(\lambda_1,\lambda_2,\ldots)$.
Various quantities depend on them and their dependence is expressed like,
$f=f(\bm{\lambda})$, $f(x)=f(x;\bm{\lambda})$.
(We sometimes omit writing $\bm{\lambda}$-dependence, when it does not
cause confusion.)

The shape invariant condition is the following \cite{os13,os14,os24}:
\begin{equation}
  \mathcal{A}(\bm{\lambda})\mathcal{A}(\bm{\lambda})^{\dagger}
  =\kappa\mathcal{A}(\bm{\lambda}+\bm{\delta})^{\dagger}
  \mathcal{A}(\bm{\lambda}+\bm{\delta})+\mathcal{E}_1(\bm{\lambda}),
  \label{shapeinv}
\end{equation}
where $\kappa$ is a real positive constant and $\bm{\delta}$ is the
shift of the parameters.
This condition combined with the Crum's theorem allows the wavefunction
$\phi_n(x)$ and energy eigenvalue $\mathcal{E}_n$ of the excited states to be
expressed in terms of the ground state wavefunction $\phi_0(x)$ and the first
excited state energy eigenvalue $\mathcal{E}_1$ with shifted parameters.
As a consequence of the shape invariance, we have
\begin{equation}
  \mathcal{A}(\bm{\lambda})\phi_n(x;\bm{\lambda})
  =f_n(\bm{\lambda})\phi_{n-1}(x;\bm{\lambda}+\bm{\delta}),\quad
  \mathcal{A}(\bm{\lambda})^{\dagger}\phi_{n-1}(x;\bm{\lambda}+\bm{\delta})
  =b_{n-1}(\bm{\lambda})\phi_n(x;\bm{\lambda}),
  \label{Aphi=,Adphi=}
\end{equation}
where $f_n(\bm{\lambda})$ and $b_{n-1}(\bm{\lambda})$ are some constants
satisfying
$f_n(\bm{\lambda})b_{n-1}(\bm{\lambda})=\mathcal{E}_n(\bm{\lambda})$.
These relations can be rewritten as the forward and backward shift relations:
\begin{equation}
  \mathcal{F}(\bm{\lambda})\check{P}_n(x;\bm{\lambda})
  =f_n(\bm{\lambda})\check{P}_{n-1}(x;\bm{\lambda}+\bm{\delta}),\quad
  \mathcal{B}(\bm{\lambda})\check{P}_{n-1}(x;\bm{\lambda}+\bm{\delta})
  =b_{n-1}(\bm{\lambda})\check{P}_n(x;\bm{\lambda}).
  \label{FP=,BP=}
\end{equation}
Here the forward and backward shift operators $\mathcal{F}(\bm{\lambda})$ and
$\mathcal{B}(\bm{\lambda})$ are defined by
\begin{align}
  &\mathcal{F}(\bm{\lambda})\eqdef
  \phi_0(x;\bm{\lambda}+\bm{\delta})^{-1}\circ
  \mathcal{A}(\bm{\lambda})\circ\phi_0(x;\bm{\lambda})
  =i\varphi(x)^{-1}(e^{\frac{\gamma}{2}p}-e^{-\frac{\gamma}{2}p}),
  \label{Fdef}\\
  &\mathcal{B}(\bm{\lambda})\eqdef
  \phi_0(x;\bm{\lambda})^{-1}\circ
  \mathcal{A}(\bm{\lambda})^{\dagger}
  \circ\phi_0(x;\bm{\lambda}+\bm{\delta})
  =-i\bigl(V(x;\bm{\lambda})e^{\frac{\gamma}{2}p}
  -V^*(x;\bm{\lambda})e^{-\frac{\gamma}{2}p}\bigr)\varphi(x),
  \label{Bdef}
\end{align}
where $\varphi(x)$ is an auxiliary function
($\varphi(x)\propto\eta(x-i\frac{\gamma}{2})-\eta(x+i\frac{\gamma}{2})$).
The difference operator $\widetilde{\mathcal{H}}$
acting on the polynomial eigenfunctions is square root free:
\begin{align}
  &\widetilde{\mathcal{H}}\eqdef
  \phi_0(x)^{-1}\circ\mathcal{H}
  \circ\phi_0(x)
  =\mathcal{B}\mathcal{F}\n
  &\phantom{\widetilde{\mathcal{H}}_{\ell}}
  =V(x)(e^{\gamma p}-1)
  +V^*(x)(e^{-\gamma p}-1),\\
  &\widetilde{\mathcal{H}}\check{P}_n(x)
  =\mathcal{E}_n\check{P}_n(x)\ \ (n=0,1,2,\ldots).
  \label{HtP=EP}
\end{align}

\medskip

By the Darboux transformation, we can deform idQM systems keeping their
exact solvability.
The multi-step Darboux transformations with virtual state wavefunctions as
seed solutions give iso-spectral deformations and the case-(1) multi-indexed
orthogonal polynomials are obtained \cite{os27,idQMcH}.
The virtual state wavefunctions are obtained by using the twist operation.
The twist operation $\mathfrak{t}$ is a map for parameters $\bm{\lambda}$
and gives a linear relation between two Hamiltonians:
\begin{equation}
  \mathcal{H}(\bm{\lambda})
  =\alpha(\bm{\lambda})\mathcal{H}\bigl(\mathfrak{t}(\bm{\lambda})\bigr)
  +\alpha'(\bm{\lambda}),
\end{equation}
where $\alpha$ and $\alpha'$ are constants.
The constant $\tilde{\bm{\delta}}$ is introduced as
$\mathfrak{t}(\bm{\lambda}+\beta\bm{\delta})=\mathfrak{t}(\bm{\lambda})
+\beta\tilde{\bm{\delta}}$ ($\forall\beta\in\mathbb{R}$).
There are two types of twist operations (type $\I$ and $\II$) for cH, W and
AW systems, and one type of twist operation for MP system.
The virtual state wavefunctions $\tilde{\phi}_{\text{v}}(x)$ are obtained from
the eigenfunction $\phi_n(x)$ \eqref{phin} as follows:
\begin{align}
  &\tilde{\phi}_{\text{v}}(x;\bm{\lambda})
  \eqdef\phi_{\text{v}}\bigl(x;\mathfrak{t}(\bm{\lambda})\bigr)
  =\phi_0\bigl(x;\mathfrak{t}(\bm{\lambda})\bigr)
  \check{P}_{\text{v}}\bigl(x;\mathfrak{t}(\bm{\lambda})\bigr),\n
  &\tilde{\xi}_{\text{v}}(x;\bm{\lambda})
  \eqdef\xi_{\text{v}}\bigl(\eta(x);\bm{\lambda}\bigr)
  \eqdef\check{P}_{\text{v}}\bigl(x;\mathfrak{t}(\bm{\lambda})\bigr)
  =P_{\text{v}}\bigl(\eta(x);\mathfrak{t}(\bm{\lambda})\bigr),
\end{align}
which satisfies the Schr\"odinger equation
$\mathcal{H}\tilde{\phi}_{\text{v}}(x)
=\tilde{\mathcal{E}}_{\text{v}}\tilde{\phi}_{\text{v}}(x)$ with the virtual
state energy $\tilde{\mathcal{E}}_{\text{v}}$,
\begin{equation}
  \tilde{\mathcal{E}}_{\text{v}}(\bm{\lambda})
  =\alpha(\bm{\lambda})
  \mathcal{E}_{\text{v}}\bigl(\mathfrak{t}(\bm{\lambda})\bigr)
  +\alpha'(\bm{\lambda}).
\end{equation}
The Hamiltonian is deformed as
$\mathcal{H}\to\mathcal{H}_{d_1}\to\mathcal{H}_{d_1d_2}\to\cdots\to
\mathcal{H}_{d_1\ldots d_s}\to\cdots\to
\mathcal{H}_{d_1\ldots d_M}=\mathcal{H}_{\mathcal{D}}$
by $M$-step Darboux transformations with virtual state wavefunctions as seed
solutions. Here the index set $\mathcal{D}=\{d_1,\ldots,d_M\}$
($d_j$: mutually distinct) labels the virtual state wavefunctions used in the
transformations.
Exactly speaking, $\mathcal{D}$ is an ordered set.
For cH, W and AW systems, there are two types of virtual states
(type $\I$ and $\II$) and $\mathcal{D}$ is
$\mathcal{D}=\{d_1,\ldots,d_M\}=\{d^{\I}_1,\ldots,d^{\I}_{M_{\I}},
d^{\II}_1,\ldots,d^{\II}_{M_{\II}}\}$ ($M=M_{\I}+M_{\II}$,
$d^{\I}_j$ : mutually distinct,
$d^{\II}_j$ : mutually distinct).
Various quantities of the deformed systems are denoted as
$\mathcal{H}_{\mathcal{D}}$, $\phi_{\mathcal{D}\,n}$,
$\mathcal{A}_{\mathcal{D}}$, etc.

The Schr\"odinger equation of the deformed system is
\begin{equation}
  \mathcal{H}_{\mathcal{D}}\phi_{\mathcal{D}\,n}(x)
  =\mathcal{E}_n\phi_{\mathcal{D}\,n}(x)\ \ (n=0,1,2,\ldots).
  \label{HphiDn=}
\end{equation}
The deformed Hamiltonian $\mathcal{H}_{\mathcal{D}}$ and eigenfunctions
$\phi_{\mathcal{D}\,n}(x)$ are given by
\begin{align}
  &\mathcal{H}_{\mathcal{D}}
  \eqdef\sqrt{V_{\mathcal{D}}(x)}\,e^{\gamma p}\sqrt{V_{\mathcal{D}}^*(x)}
  +\!\sqrt{V_{\mathcal{D}}^*(x)}\,e^{-\gamma p}\sqrt{V_{\mathcal{D}}(x)}
  -V_{\mathcal{D}}(x)-V_{\mathcal{D}}^*(x)
  =\mathcal{A}_{\mathcal{D}}^{\dagger}\mathcal{A}_{\mathcal{D}},
  \label{HD}\\
  &\mathcal{A}_{\mathcal{D}}
  \eqdef i\bigl(e^{\frac{\gamma}{2}p}\sqrt{V_{\mathcal{D}}^*(x)}
  -e^{-\frac{\gamma}{2}p}\sqrt{V_{\mathcal{D}}(x)}\,\bigr),\quad
  \mathcal{A}_{\mathcal{D}}^{\dagger}
  \eqdef-i\bigl(\sqrt{V_{\mathcal{D}}(x)}\,e^{\frac{\gamma}{2}p}
  -\sqrt{V_{\mathcal{D}}^*(x)}\,e^{-\frac{\gamma}{2}p}\bigr),\\
  &V_{\mathcal{D}}(x;\bm{\lambda})
  \eqdef V(x;\bm{\lambda}')\,
  \frac{\check{\Xi}_{\mathcal{D}}(x+i\frac{\gamma}{2};\bm{\lambda})}
  {\check{\Xi}_{\mathcal{D}}(x-i\frac{\gamma}{2};\bm{\lambda})}
  \frac{\check{\Xi}_{\mathcal{D}}(x-i\gamma;\bm{\lambda}+\bm{\delta})}
  {\check{\Xi}_{\mathcal{D}}(x;\bm{\lambda}+\bm{\delta})},
  \label{VD}\\
  &\phi_{\mathcal{D}\,n}(x)\eqdef A
  \psi_{\mathcal{D}}(x)\check{P}_{\mathcal{D},n}(x),\quad
  \psi_{\mathcal{D}}(x;\bm{\lambda})\eqdef
  \frac{\phi_0(x;\bm{\lambda}')}
  {\sqrt{\check{\Xi}_{\mathcal{D}}(x-i\frac{\gamma}{2};\bm{\lambda})
  \check{\Xi}_{\mathcal{D}}(x+i\frac{\gamma}{2};\bm{\lambda})}},
  \label{phiDn}
\end{align}
where $A$ and $\bm{\lambda}'$ are
\begin{align}
  A&=\left\{
  \begin{array}{ll}
  \kappa^{-\frac14M(M+1)}\alpha(\bm{\lambda}')^{\frac12M}
  &:\text{MP}\\[2pt]
  \kappa^{-\frac14M_{\I}(M_{\I}+1)-\frac14M_{\II}(M_{\II}+1)
  +\frac52M_{\I}M_{\II}}
  \alpha^{\I}(\bm{\lambda}')^{\frac12M_{\I}}
  \alpha^{\II}(\bm{\lambda}')^{\frac12M_{\II}}
  &:\text{cH,W,AW}
  \end{array}\right.,
  \label{A:phiDn}\\
  \bm{\lambda}'&=\left\{
  \begin{array}{ll}
  \bm{\lambda}+M\tilde{\bm{\delta}}&:\text{MP}\\
  \bm{\lambda}^{[M_{\I},M_{\II}]}\eqdef
  \bm{\lambda}+M_{\I}\tilde{\bm{\delta}}^{\I}+M_{\II}\tilde{\bm{\delta}}^{\II}
  &:\text{cH,W,AW}
  \end{array}\right..
  \label{la'}
\end{align}
Note that $A=1$ for MP, cH and W.
Here $\check{\Xi}_{\mathcal{D}}(x)$ and $\check{P}_{\mathcal{D},n}(x)$ are
polynomials in $\eta(x)$,
\begin{equation}
  \check{\Xi}_{\mathcal{D}}(x)\eqdef
  \Xi_{\mathcal{D}}\bigl(\eta(x)\bigr),\quad
  \check{P}_{\mathcal{D},n}(x)\eqdef
  P_{\mathcal{D},n}\bigl(\eta(x)\bigr),
  \label{XiP_poly}
\end{equation}
and their explicit forms are given in \cite{os27,idQMcH}.
The denominator polynomial $\Xi_{\mathcal{D}}(\eta)$ and
the multi-indexed polynomial $P_{\mathcal{D},n}(\eta)$ are
polynomials in $\eta$ and their degrees are $\ell_{\mathcal{D}}$ and
$\ell_{\mathcal{D}}+n$, respectively.
Here $\ell_{\mathcal{D}}$ is
\begin{equation}
  \ell_{\mathcal{D}}\eqdef\sum_{j=1}^Md_j-\tfrac12M(M-1)+\left\{
  \begin{array}{ll}
  0&:\text{MP}\\
  2M_{\I}M_{\II}&:\text{cH,W,AW}
  \end{array}\right..
  \label{lD}
\end{equation}
The deformed Hamiltonian $\mathcal{H}_{\mathcal{D}}$ is
hermitian, if the following condition is satisfied \cite{os27,idQMcH}:
\begin{equation}
  \text{The denominator polynomial $\check{\Xi}_{\mathcal{D}}(x)$
  has no zero in $D_{\gamma}$ \eqref{Dgamma}.}
  \label{nozero}
\end{equation}
The eigenfunctions $\phi_{\mathcal{D}\,n}(x)$ are orthogonal each other,
which gives the orthogonality relation of $\check{P}_{\mathcal{D},n}(x)$ :
\begin{align}
  &\int_{x_1}^{x_2}\!\!dx\,\psi_{\mathcal{D}}(x)^2
  \check{P}_{\mathcal{D},n}(x)
  \check{P}_{\mathcal{D},m}(x)
  =h_{\mathcal{D},n}\delta_{nm}
  \ \ (n,m=0,1,2,\ldots),
  \label{orthocPDn}\\
  &h_{\mathcal{D},n}=A^{-2}h_n\times\left\{
  \begin{array}{ll}
  \prod_{j=1}^M(\mathcal{E}_n-\tilde{\mathcal{E}}_{d_j})&:\text{MP}\\[6pt]
  \prod_{j=1}^{M_{\I}}(\mathcal{E}_n
  -\tilde{\mathcal{E}}^{\I}_{d^{\I}_j})\cdot
  \prod_{j=1}^{M_{\II}}(\mathcal{E}_n
  -\tilde{\mathcal{E}}^{\II}_{d^{\II}_j})&:\text{cH,W,AW}
  \end{array}\right.,
  \label{hDn}
\end{align}
where $A$ is given by \eqref{A:phiDn}.
The multi-indexed orthogonal polynomial $P_{\mathcal{D},n}(\eta)$
has $n$ zeros in the physical region ($\eta(x_1)<\eta<\eta(x_2)$ for MP, cH, W,
$\eta(x_2)<\eta<\eta(x_1)$ for AW),
which interlace the $n+1$ zeros of
$P_{\mathcal{D},n+1}(\eta)$ in the physical region,
and $\ell_{\mathcal{D}}$ zeros in the unphysical region
($\eta\in\mathbb{C}\backslash\{\text{physical region}$ $\text{of $\eta$}\}$).

The shape invariance of the original system is inherited by the deformed
systems.
By the argument of \cite{os27}, the Hamiltonian
$\mathcal{H}_{\mathcal{D}}(\bm{\lambda})$ is shape invariant:
\begin{equation}
  \mathcal{A}_{\mathcal{D}}(\bm{\lambda})
  \mathcal{A}_{\mathcal{D}}(\bm{\lambda})^{\dagger}
  =\kappa\mathcal{A}_{\mathcal{D}}(\bm{\lambda}+\bm{\delta})^{\dagger}
  \mathcal{A}_{\mathcal{D}}(\bm{\lambda}+\bm{\delta})
  +\mathcal{E}_1(\bm{\lambda}).
  \label{shapeinvD}
\end{equation}
As a consequence of the shape invariance, the actions of
$\mathcal{A}_{\mathcal{D}}(\bm{\lambda})$ and
$\mathcal{A}_{\mathcal{D}}(\bm{\lambda})^{\dagger}$ on the eigenfunctions
$\phi_{\mathcal{D}\,n}(x;\bm{\lambda})$ are
\begin{align}
  \mathcal{A}_{\mathcal{D}}(\bm{\lambda})
  \phi_{\mathcal{D}\,n}(x;\bm{\lambda})
  &=\kappa^{\frac{M}{2}}f_n(\bm{\lambda})
  \phi_{\mathcal{D}\,n-1}(x;\bm{\lambda}+\bm{\delta}),\n
  \mathcal{A}_{\mathcal{D}}(\bm{\lambda})^{\dagger}
  \phi_{\mathcal{D}\,n-1}(x;\bm{\lambda}+\bm{\delta})
  &=\kappa^{-\frac{M}{2}}b_{n-1}(\bm{\lambda})
  \phi_{\mathcal{D}\,n}(x;\bm{\lambda}).
  \label{ADphiDn=,ADdphiDn=}
\end{align}
The forward and backward shift operators are defined by
\begin{align}
  \mathcal{F}_{\mathcal{D}}(\bm{\lambda})&\eqdef
  \psi_{\mathcal{D}}\,(x;\bm{\lambda}+\bm{\delta})^{-1}\circ
  \mathcal{A}_{\mathcal{D}}(\bm{\lambda})\circ
  \psi_{\mathcal{D}}\,(x;\bm{\lambda})\n
  &=\frac{i}{\varphi(x)\check{\Xi}_{\mathcal{D}}(x;\bm{\lambda})}
  \Bigl(\check{\Xi}_{\mathcal{D}}(x+i\tfrac{\gamma}{2};
  \bm{\lambda}+\bm{\delta})e^{\frac{\gamma}{2}p}
  -\check{\Xi}_{\mathcal{D}}(x-i\tfrac{\gamma}{2};
  \bm{\lambda}+\bm{\delta})e^{-\frac{\gamma}{2}p}\Bigr),
  \label{calFD}\\
  \mathcal{B}_{\mathcal{D}}(\bm{\lambda})&\eqdef
  \psi_{\mathcal{D}}\,(x;\bm{\lambda})^{-1}\circ
  \mathcal{A}_{\mathcal{D}}(\bm{\lambda})^{\dagger}\circ
  \psi_{\mathcal{D}}\,(x;\bm{\lambda}+\bm{\delta})
  \label{calBD}\\
  &=\frac{-i}{\check{\Xi}_{\mathcal{D}}(x;\bm{\lambda}+\bm{\delta})}
  \Bigl(V(x;\bm{\lambda}')
  \check{\Xi}_{\mathcal{D}}(x+i\tfrac{\gamma}{2};\bm{\lambda})
  e^{\frac{\gamma}{2}p}
  -V^*(x;\bm{\lambda}')
  \check{\Xi}_{\mathcal{D}}(x-i\tfrac{\gamma}{2};\bm{\lambda})
  e^{-\frac{\gamma}{2}p}\Bigr)\varphi(x),
  \nonumber
\end{align}
($\bm{\lambda'}$ is given by \eqref{la'})
and their actions on $\check{P}_{\mathcal{D},n}(x;\bm{\lambda})$ are
\begin{align}
  \mathcal{F}_{\mathcal{D}}(\bm{\lambda})
  \check{P}_{\mathcal{D},n}(x;\bm{\lambda})
  &=f_n(\bm{\lambda})
  \check{P}_{\mathcal{D},n-1}(x;\bm{\lambda}+\bm{\delta}),\n
  \mathcal{B}_{\mathcal{D}}(\bm{\lambda})
  \check{P}_{\mathcal{D},n-1}(x;\bm{\lambda}+\bm{\delta})
  &=b_{n-1}(\bm{\lambda})
  \check{P}_{\mathcal{D},n}(x;\bm{\lambda}).\!
  \label{FDPDn=,BDPDn=}
\end{align}
The similarity transformed Hamiltonian is square root free:
\begin{align}
  \widetilde{\mathcal{H}}_{\mathcal{D}}(\bm{\lambda})
  &\eqdef\psi_{\mathcal{D}}(x;\bm{\lambda})^{-1}\circ
  \mathcal{H}_{\mathcal{D}}(\bm{\lambda})\circ
  \psi_{\mathcal{D}}(x;\bm{\lambda})
  =\mathcal{B}_{\mathcal{D}}(\bm{\lambda})
  \mathcal{F}_{\mathcal{D}}(\bm{\lambda})\n
  &=V(x;\bm{\lambda}')\,
  \frac{\check{\Xi}_{\mathcal{D}}(x+i\frac{\gamma}{2};\bm{\lambda})}
  {\check{\Xi}_{\mathcal{D}}(x-i\frac{\gamma}{2};\bm{\lambda})}
  \biggl(e^{\gamma p}
  -\frac{\check{\Xi}_{\mathcal{D}}(x-i\gamma;\bm{\lambda}+\bm{\delta})}
  {\check{\Xi}_{\mathcal{D}}(x;\bm{\lambda}+\bm{\delta})}\biggr)\n
  &\quad+V^*(x;\bm{\lambda}')\,
  \frac{\check{\Xi}_{\mathcal{D}}(x-i\frac{\gamma}{2};\bm{\lambda})}
  {\check{\Xi}_{\mathcal{D}}(x+i\frac{\gamma}{2};\bm{\lambda})}
  \biggl(e^{-\gamma p}
  -\frac{\check{\Xi}_{\mathcal{D}}(x+i\gamma;\bm{\lambda}+\bm{\delta})}
  {\check{\Xi}_{\mathcal{D}}(x;\bm{\lambda}+\bm{\delta})}\biggr).
  \label{tHD}
\end{align}
By defining $V'_{\mathcal{D}}(x)$ as
\begin{equation}
  V'_{\mathcal{D}}(x;\bm{\lambda})\eqdef V(x;\bm{\lambda}')\,
  \frac{\check{\Xi}_{\mathcal{D}}(x+i\frac{\gamma}{2};\bm{\lambda})}
  {\check{\Xi}_{\mathcal{D}}(x-i\frac{\gamma}{2};\bm{\lambda})},
  \label{VpD}
\end{equation}
it is written as
\begin{equation}
  \widetilde{\mathcal{H}}_{\mathcal{D}}
  =V'_{\mathcal{D}}(x)e^{\gamma p}
  +V^{\prime\,*}_{\mathcal{D}}(x)e^{-\gamma p}
  -V_{\mathcal{D}}(x)-V^*_{\mathcal{D}}(x).
  \label{tHD2}
\end{equation}
The multi-indexed orthogonal polynomials
$\check{P}_{\mathcal{D},n}(x)$ are its eigenpolynomials:
\begin{equation}
  \widetilde{\mathcal{H}}_{\mathcal{D}}\check{P}_{\mathcal{D},n}(x)
  =\mathcal{E}_n\check{P}_{\mathcal{D},n}(x)\ \ (n=0,1,2,\ldots).
  \label{tHPDn=}
\end{equation}

\section{Some Properties of the Multi-indexed Meixner-Pol\-la\-czek Polynomials}
\label{app:MP:prop}

We present some properties of the multi-indexed Meixner-Pollaczek polynomials
\cite{idQMcH}.

\noindent
$\bullet$ coefficients of the highest degree terms :
\begin{align}
  \Xi_{\mathcal{D}}(\eta;\bm{\lambda})
  &=c_{\mathcal{D}}^{\Xi}(\bm{\lambda})\eta^{\ell_{\mathcal{D}}}
  +(\text{lower order terms}),\n
  c_{\mathcal{D}}^{\Xi}(\bm{\lambda})
  &=\prod_{j=1}^Mc_{d_j}\bigl(\mathfrak{t}(\bm{\lambda})\bigr)\cdot
  \prod_{1\leq j<k\leq M}(d_k-d_j),
  \label{MP:cXiD}\\
  P_{\mathcal{D}}(\eta;\bm{\lambda})
  &=c_{\mathcal{D},n}^{P}(\bm{\lambda})\eta^{\ell_{\mathcal{D}}+n}
  +(\text{lower order terms}),\n
  c_{\mathcal{D},n}^{P}(\bm{\lambda})
  &=c_{\mathcal{D}}^{\Xi}(\bm{\lambda})c_n(\bm{\lambda})
  \prod_{j=1}^M(-2a-n+d_j+1).
  \label{MP:cPDn}
\end{align}

\noindent
$\bullet$ $\check{P}_{\mathcal{D},0}(x;\bm{\lambda})$ vs
$\check{\Xi}_{\mathcal{D}}(x;\bm{\lambda})$ :
\begin{equation}
  \check{P}_{\mathcal{D},0}(x;\bm{\lambda})
  =A\,\check{\Xi}_{\mathcal{D}}(x;\bm{\lambda}+\bm{\delta}),\quad
  A=\prod_{j=1}^M(-2a+d_j+1).
  \label{MP:PD0=A.XiD}
\end{equation}

\noindent
$\bullet$ $d_j=0$ case :
\begin{align}
  &\check{P}_{\mathcal{D},n}(x;\bm{\lambda})\Bigm|_{d_M=0}
  =A\,\check{P}_{\mathcal{D}',n}(x;\bm{\lambda}+\tilde{\bm{\delta}}),\quad
  \mathcal{D}'=\{d_1-1,\ldots,d_{M-1}-1\},\n[4pt]
  &\quad A=(-1)^M(2a+n-1)(2\sin\phi)^{M-1}.
  \label{MP:dM=0}
\end{align}

\noindent
$\bullet$ values at special points :\\
Let $x_0$ and $\eta_0$ be
\begin{equation}
  x_0\eqdef -i\gamma(a-\tfrac12M),\quad
  \eta_0\eqdef\eta(x_0).
\end{equation}
Note that, as coordinates $x$ and $\eta$, these values $x_0$ and $\eta_0$ are
unphysical (they are imaginary).
The multi-indexed polynomials $P_{\mathcal{D},n}(\eta)$ take `simple' values
at these `unphysical' values $\eta_0$:
\begin{equation}
  P_{\mathcal{D},n}(\eta_0;\bm{\lambda})
  =c^P_{\mathcal{D},n}(\bm{\lambda})
  e^{i\phi(\ell_{\mathcal{D}}-n)}
  \frac{(2a)_n}{(2\sin\phi)^{\ell_{\mathcal{D}}+n}}
  \prod_{j=1}^M\frac{(1-2a)_{d_j}}{(1-2a)_{j-1}}\cdot
  \prod_{j=1}^M\frac{d_j+1-2a}{d_j+1-n-2a},
  \label{MP:PDn(eta0)}
\end{equation}
where we have assumed $0\leq d_1<\cdots<d_M$.

\section{Some Properties of the Multi-indexed Continuous Hahn Polynomials}
\label{app:cH:prop}

We present some properties of the multi-indexed continuous Hahn polynomials
\cite{idQMcH}.

\noindent
$\bullet$ coefficients of the highest degree terms :
\begin{align}
  \Xi_{\mathcal{D}}(\eta;\bm{\lambda})
  &=c_{\mathcal{D}}^{\Xi}(\bm{\lambda})\eta^{\ell_{\mathcal{D}}}
  +(\text{lower order terms}),\n
  c_{\mathcal{D}}^{\Xi}(\bm{\lambda})
  &=\prod_{j=1}^{M_{\I}}c_{d^{\I}_j}
  \bigl(\mathfrak{t}^{\I}(\bm{\lambda})\bigr)\cdot
  \prod_{j=1}^{M_{\II}}c_{d^{\II}_j}
  \bigl(\mathfrak{t}^{\II}(\bm{\lambda})\bigr)\cdot
  \prod_{1\leq j<k\leq M_{\I}}(d^{\I}_k-d^{\I}_j)\cdot
  \prod_{1\leq j<k\leq M_{\II}}(d^{\II}_k-d^{\II}_j)\n
  &\quad\times
  \prod_{j=1}^{M_{\I}}\prod_{k=1}^{M_{\II}}
  (-a_2-a_2^*-d^{\I}_j+a_1+a_1^*+d^{\II}_k),
  \label{cXiD}\\
  P_{\mathcal{D}}(\eta;\bm{\lambda})
  &=c_{\mathcal{D},n}^{P}(\bm{\lambda})\eta^{\ell_{\mathcal{D}}+n}
  +(\text{lower order terms}),\n
  c_{\mathcal{D},n}^{P}(\bm{\lambda})
  &=c_{\mathcal{D}}^{\Xi}(\bm{\lambda})c_n(\bm{\lambda})
  \prod_{j=1}^{M_{\I}}(-a_1-a_1^*-n+d^{\I}_j+1)\cdot
  \prod_{j=1}^{M_{\II}}(-a_2-a_2^*-n+d^{\II}_j+1).
  \label{cPDn}
\end{align}

\noindent
$\bullet$ $\check{P}_{\mathcal{D},0}(x;\bm{\lambda})$ vs
$\check{\Xi}_{\mathcal{D}}(x;\bm{\lambda})$ :
\begin{align}
  &\check{P}_{\mathcal{D},0}(x;\bm{\lambda})
  =A\,\check{\Xi}_{\mathcal{D}}(x;\bm{\lambda}+\bm{\delta}),\n
  &\quad A=\prod_{j=1}^{M_{\I}}(-a_1-a_1^*+d^{\I}_j+1)
  \cdot\prod_{j=1}^{M_{\II}}(-a_2-a_2^*+d^{\II}_j+1).
  \label{PD0=A.XiD}
\end{align}

\noindent
$\bullet$ $d_j=0$ case :
\begin{align}
  &\check{P}_{\mathcal{D},n}(x;\bm{\lambda})\Bigm|_{d^{\I}_{M_{\I}}=0}
  =A\,\check{P}_{\mathcal{D}',n}(x;\bm{\lambda}+\tilde{\bm{\delta}}^{\I}),\n
  &\quad\mathcal{D}'=\{d^{\I}_1-1,\ldots,d^{\I}_{M_{\I}-1}-1,
  d^{\II}_1+1,\ldots,d^{\II}_{M_{\II}}+1\},\n
  &\quad A=(-1)^{M_{\I}}(a_1+a_1^*+n-1)
  \prod_{j=1}^{M_{\I}-1}(-a_1-a_1^*+a_2+a_2^*+d^{\I}_j+1)\cdot
  \prod_{j=1}^{M_{\II}}(d^{\II}_j+1),
  \label{dIM1=0}\\
  &\check{P}_{\mathcal{D},n}(x;\bm{\lambda})\Bigm|_{d^{\II}_{M_{\II}}=0}
  =B\,\check{P}_{\mathcal{D}',n}(x;\bm{\lambda}+\tilde{\bm{\delta}}^{\II}),\n
  &\quad\mathcal{D}'=\{d^{\I}_1+1,\ldots,d^{\I}_{M_{\I}}+1,
  d^{\II}_1-1,\ldots,d^{\II}_{M_{\II}-1}-1\},\n
  &\quad B=(-1)^M(a_2+a_2^*+n-1)
  \prod_{j=1}^{M_{\II}-1}(-a_2-a_2^*+a_1+a_1^*+d^{\II}_j+1)\cdot
  \prod_{j=1}^{M_{\I}}(d^{\I}_j+1).
  \label{dIIM2=0}
\end{align}

\noindent
$\bullet$ values at special points :\\
Let $x_0$ and $\eta_0$ be
\begin{equation}
  x_0\eqdef -i\gamma\bigl(a_2^*+\tfrac12(M_{\I}-M_{\II})\bigr),\quad
  \eta_0\eqdef\eta(x_0).
\end{equation}
Note that, as coordinates $x$ and $\eta$, these values $x_0$ and $\eta_0$ are
unphysical (they are imaginary).
The multi-indexed polynomials $P_{\mathcal{D},n}(\eta)$ take `simple' values
at these `unphysical' values $\eta_0$:
\begin{align}
  &\quad P_{\mathcal{D},n}(\eta_0;\bm{\lambda})\n
  &=(-i)^{\ell_{\mathcal{D}}+n}c^P_{\mathcal{D},n}(\bm{\lambda})
  (-1)^{\sum_{j=1}^{M_{\II}}d^{\II}_j-\frac12M_{\II}(M_{\II}-1)}\n
  &\quad\times
  \prod_{j=1}^{M_{\I}}\frac{(a_2^*-a_1^*+1,a_2+a_2^*)_{d^{\I}_j}}
  {(a_2+a_2^*-a_1-a_1^*+d^{\I}_j+1)_{d^{\I}_j}}\cdot
  \frac{\prod\limits_{1\leq j<k\leq M_{\I}}
  (a_2+a_2^*-a_1-a_1^*+d^{\I}_j+d^{\I}_k+1)}
  {\prod\limits_{j=1}^{M_{\I}}(a_2^*-a_1^*+1,a_2+a_2^*)_{j-1}}\n
  &\quad\times
  \prod_{j=1}^{M_{\II}}\frac{(a_1^*-a_2^*+1,1-a_2-a_2^*)_{d^{\II}_j}}
  {(a_1+a_1^*-a_2-a_2^*+d^{\II}_j+1)_{d^{\II}_j}}\cdot
  \frac{\prod\limits_{1\leq j<k\leq M_{\II}}
  (a_1+a_1^*-a_2-a_2^*+d^{\II}_j+d^{\II}_k+1)}
  {\prod\limits_{j=1}^{M_{\II}}(a_1^*-a_2^*+1,1-a_2-a_2^*)_{j-1}}\n
  &\quad\times
  \prod_{j=1}^{M_{\I}}\prod_{k=1}^{M_{\II}}
  \frac{(a_2^*-a_1^*+j-k)(a_2+a_2^*+j-k)}
  {a_2+a_2^*-a_1-a_1^*+d^{\I}_j-d^{\II}_k}\n
  &\quad\times
  \frac{(a_1+a_2^*,a_2+a_2^*)_n}{(a_1+a_1^*+a_2+a_2^*+n-1)_n}
  \prod_{j=1}^{M_{\I}}\frac{a_2+a_2^*+d^{\I}_j+n}{a_2+a_2^*+j-1}\cdot
  \prod_{j=1}^{M_{\II}}\frac{d^{\II}_j+1-a_2-a_2^*}{d^{\II}_j+1-n-a_2-a_2^*},
   \label{PDn(eta0)}
\end{align}
where we have assumed $0\leq d^{\I}_1<\cdots<d^{\I}_{M_{\I}}$
and $0\leq d^{\II}_1<\cdots<d^{\II}_{M_{\II}}$.
We remark that a similar formula can be obtained by interchanging
$a_1\leftrightarrow a_2$ and $\I\leftrightarrow\II$.

\section{More Examples for \S\,\ref{sec:rr}}
\label{app:ex}

We present more examples for \S\,\ref{sec:rr}.
Unlike in \S\,\ref{sec:ex}, the examples presented here do not satisfy the
condition \eqref{nozero}.
Namely, the multi-indexed polynomials $P_{\mathcal{D},n}(\eta)$ are not
orthogonal polynomials.
Except for Ex.1 in \S\,\ref{app:ex_MP},
we present only $r_{n,k}^{X,\mathcal{D}}$ ($1\leq k\leq L$), because
$r_{n,k}^{X,\mathcal{D}}$ ($-L\leq k\leq 0$) are obtained by
\eqref{prop_rnk}--\eqref{rn0}.

An explicit form of $A$ in \eqref{A_Ex1_cH} is also presented.

\subsection{Examples for \S\,\ref{sec:ex_MP}}
\label{app:ex_MP}

\noindent
\underline{Ex.1} $\mathcal{D}=\{1\}$, $Y(\eta)=1$
($\Rightarrow\ell_{\mathcal{D}}=1,\,X(\eta)=X_{\text{min}}(\eta),\,L=2$):
5-term recurrence relations
\begin{align*}
  X(\eta)&=\eta\bigl(\sin\phi\cdot\eta+2(1-a)\cos\phi\bigr),\\
  r_{n,2}^{X,\mathcal{D}}&=\frac{(n+1)_2}{4\sin\phi}\frac{2a+n-2}{2a+n},\quad
  r_{n,-2}^{X,\mathcal{D}}=\frac{(2a+n-2)_2}{4\sin\phi},\\
  r_{n,1}^{X,\mathcal{D}}&=-(n+1)(2a+n-2)\cot\phi,\quad
  r_{n,-1}^{X,\mathcal{D}}=-(2a+n-2)_2\,\cot\phi,\\
  r_{n,0}^{X,\mathcal{D}}&=
  \frac{a(6a+10n-7)+3n^2-7n+1}{2\sin\phi}
  -\frac14(2a+2n-1)(6a+2n-5)\sin\phi.
\end{align*}
The polynomial $I(z)$ \eqref{Iz} is
\begin{align*}
  I(z)&=-8\sin\phi\Bigl(
  (2+\cos 2\phi)z^2
  +2\bigl(6a-4+(4a-3)\cos 2\phi\bigr)\sin\phi\cdot z\\
  &\quad
  +\bigl(12a(a-1)-1+(2a-1)(6a-5)\cos 2\phi\bigr)\sin^2\phi\Bigr).
\end{align*}

\noindent
\underline{Ex.2} $\mathcal{D}=\{1\}$, $Y(\eta)=\eta$
($\Rightarrow\ell_{\mathcal{D}}=1,\,L=3$):
7-term recurrence relations
\begin{align*}
  X(\eta)&=\frac{\eta}{6}\Bigl(4\sin\phi\cdot\eta^2
  +6(1-a)\cos\phi\cdot\eta+\sin\phi\Bigr),\\
  r_{n,3}^{X,\mathcal{D}}&=\frac{(n+1)_3}{12\sin^2\phi}
  \frac{2a+n-2}{2a+n+1},\quad
  r_{n,2}^{X,\mathcal{D}}=-\frac{(n+1)_2(3a+2n)\cos\phi}{4\sin^2\phi}
  \frac{2a+n-2}{2a+n},\\
  r_{n,1}^{X,\mathcal{D}}&=\frac{(n+1)(2a+n-2)}{4\sin^2\phi}
  \bigl(2a+3n+1+2(a+n)\cos 2\phi\bigr).
\end{align*}
The polynomial $I(z)$ \eqref{Iz} is
\begin{align*}
  I(z)&=-48\sin 2\phi\Bigl(
  (4+\cos 2\phi)z^3
  +3\bigl(10a-6+(3a-2)\cos 2\phi\bigr)\sin\phi\cdot z^2\\
  &\quad
  +2\bigl(30a(a-1)+4+(12a^2-15a+4)\cos 2\phi\bigr)\sin^2\phi\cdot z\\
  &\quad
  +(a-1)\bigl(20a(a+1)-3+(2a-1)(10a-3)\cos 2\phi\bigr)\sin^3\phi\Bigr).
\end{align*}

\noindent
\underline{Ex.3} $\mathcal{D}=\{1\}$, $Y(\eta)=\eta^2$
($\Rightarrow\ell_{\mathcal{D}}=1,\,L=4$):
9-term recurrence relations
\begin{align*}
  X(\eta)&=\frac{\eta}{12}\bigl(
  6\sin\phi\cdot\eta^3+8(1-a)\cos\phi\cdot\eta^2
  +3\sin\phi\cdot\eta+2(1-a)\cos\phi\bigr),\\
  r_{n,4}^{X,\mathcal{D}}&=\frac{(n+1)_4}{32\sin^3\phi}
  \frac{2a+n-2}{2a+n+2},\quad
  r_{n,3}^{X,\mathcal{D}}=-\frac{(n+1)_3(4a+3n+2)\cos\phi}{12\sin^3\phi}
  \frac{2a+n-2}{2a+n+1},\\
  r_{n,2}^{X,\mathcal{D}}&=\frac{(n+1)_2}{8\sin^3\phi}
  \frac{2a+n-2}{2a+n}\bigl(
  5a(a+1)+2(5a+1)n+4n^2+(5a^2+2a+8an+3n^2+n)\cos2\phi\bigr),\\
  r_{n,1}^{X,\mathcal{D}}&=-\frac{(n+1)(2a+n-2)\cos\phi}{4\sin^3\phi}
  \bigl(2a^2+4a+8an+5n^2+2n+1+2(a+n)^2\cos 2\phi\bigr).
\end{align*}
The polynomial $I(z)$ \eqref{Iz} is
\begin{align*}
  I(z)&=-192\sin\phi\Bigl(
  3(18+16\cos 2\phi+\cos 4\phi)z^4\\
  &\quad
  +4\bigl(18(7a-4)+4(29a-17)\cos 2\phi+(8a-5)\cos 4\phi\bigr)
  \sin\phi\cdot z^3\\
  &\quad
  +12\bigl(126a(a-1)+30+4(31a^2-33a+8)\cos 2\phi
  +(10a^2-12a+3)\cos 4\phi\bigr)\sin^2\phi\cdot z^2\\
  &\quad
  +16\bigl(3(32a^3-33a^2+13a-6)+2(54a^3-69a^2+29a-8)\cos 2\phi\\
  &\qquad
  +(a-1)(12a^2-9a+1)\cos 4\phi\bigr)\sin^3\phi\cdot z\\
  &\quad
  +\bigl(3(2a+1)(56a^3+4a^2-58a-3)
  +4(112a^4-16a^3-112a^2+16a+3)\cos 2\phi\\
  &\qquad
  +(2a-1)(56a^3-100a^2+38a+3)\cos 4\phi\bigr)
  \sin^4\phi\Bigr).
\end{align*}

\noindent
\underline{Ex.4} $\mathcal{D}=\{3\}$, $Y(\eta)=1$
($\Rightarrow\ell_{\mathcal{D}}=3,\,X(\eta)=X_{\text{min}}(\eta),\,L=4$):
9-term recurrence relations
\begin{align*}
  X(\eta)&=\frac{\eta}{12}\Bigl(
  4\sin^3\phi\cdot\eta^3
 -16(a-2)\cos\phi\sin^2\phi\cdot\eta^2\\
  &\quad
  +\bigl(12a(a-3)+29+(12a(a-4)+43)\cos 2\phi\bigr)\sin\phi\cdot\eta\\
  &\quad
  -2(a-2)\bigl((2a-1)^2+(4a(a-4)+11)\cos 2\phi\bigr)\cos\phi\Bigr),\\
  r_{n,4}^{X,\mathcal{D}}&=\frac{(n+1)_4}{48\sin\phi}
  \frac{2a+n-4}{2a+n},\quad
  r_{n,3}^{X,\mathcal{D}}=-\frac16(n+1)_3\,(2a+n-4)\cot\phi,\\
  r_{n,2}^{X,\mathcal{D}}&=\frac{(n+1)_2}{12\sin\phi}(2a+n-4)
  \bigl(6a+4n+3(2a+n-1)\cos 2\phi\bigr),\\
  r_{n,1}^{X,\mathcal{D}}&=-\frac16(n+1)(2a+n-1)(2a+n-4)\cot\phi
  \bigl(4a+5n+2+2(2a+n-2)\cos 2\phi\bigr).
\end{align*}
The polynomial $I(z)$ \eqref{Iz} is
\begin{align*}
  I(z)&=-384\sin^3\phi\Bigl(
  (18+16\cos 2\phi+\cos 4\phi)z^4\\
  &\quad
  +4\bigl(54(a-1)+4(13a-14)\cos 2\phi+(4a-5)\cos 4\phi\bigr)\sin\phi\cdot z^3\\
  &\quad
  +4\bigl(6(2a-1)(19a-27)+8(30a^2-63a+28)\cos 2\phi
  +(24a^2-60a+35)\cos 4\phi\bigr)\sin^2\phi\cdot z^2\\
  &\quad
  +16\bigl(3(2a-1)(16a^2-35a+12)+8(14a^3-42a^2+35a-8)\cos 2\phi\\
  &\qquad
  +(4a-5)(4a^2-10a+5)\cos 4\phi\bigr)\sin^3\phi\cdot z\\
  &\quad
  +3(2a-1)\bigl((2a-1)(60a^2-116a-9)+4(40a^3-116a^2+70a+5)\cos 2\phi\\
  &\qquad
  +(2a-3)(20a^2-56a+31)\cos 4\phi\bigr)\sin^4\phi\Bigr).
\end{align*}

\ignore{-----------------------
\noindent
\underline{Ex.5} $\mathcal{D}=\{1,3\}$, $Y(\eta)=1$
($\Rightarrow X(\eta)=X_{\text{min}}(\eta)$, $L=4$):
9-term recurrence relations
\begin{align*}
  X(\eta)&=\frac{\eta}{3}\Bigl(
  4\sin^4\phi\cdot\eta^3
  -8(2a-3)\cos\phi\sin^3\phi\cdot\eta^2\\
  &\quad
  +2\bigl(6a^2-18a+13+(6a^2-18a+11)\cos 2\phi\bigr)\sin^2\phi\cdot\eta\\
  &\quad
  -2(2a-3)\bigl(a^2-3a+3+(a^2-3a+1)\cos 2\phi\bigr)\sin 2\phi\Bigr),\\
  r_{n,4}^{X,\mathcal{D}}&=\frac{(n+1)_4}{12}
  \frac{(2a+n-2)(2a+n-4)}{(2a+n)(2a+n+2)},\quad
  r_{n,3}^{X,\mathcal{D}}=\frac{2(n+1)_3\cos\phi}{3}
  \frac{(2a+n-2)(2a+n-4)}{2a+n+1},\\
  r_{n,2}^{X,\mathcal{D}}&=\frac{(n+1)_2}{3}
  \frac{2a+n-4}{2a+n}
  \Bigl(a(14n-15)+4n^2-9n+3+24a^2\cos^2\phi\\
  &\qquad
  +3\bigl(2a(2n-3)+(n-1)(n-2)\bigr)\cos 2\phi\Bigr),\\
  r_{n,1}^{X,\mathcal{D}}&=-\frac{2(n+1)\cos\phi}{3}
  (2a+n-2)(2a+n-4)\bigl(4a+5n-1+2(2a+n-2)\cos 2\phi\bigr).
\end{align*}
The polynomial $I(z)$ \eqref{Iz} is
\begin{align*}
  I(z)&=-1536\sin^4\phi\Bigl(
  (18+16\cos 2\phi+\cos 4\phi)z^4\\
  &\quad
  +4\bigl(9(6a-7)+2(26a-31)\cos 2\phi+(4a-5)\cos 4\phi\bigr)\sin\phi\cdot z^3\\
  &\quad
  +4\bigl(6(38a^2-87a+43)+4(60a^2-141a+74)\cos 2\phi+(24a^2-60a+35)\cos 4\phi
  \bigr)\sin^2\phi\cdot z^2\\
  &\quad
  +16\bigl(3(a-1)(2a-1)(16a-29)+2(56a^3-192a^2+197a-59)\cos 2\phi\\
  &\qquad
  +(4a-5)(4a^2-10a+5)\cos 4\phi\bigr)\sin^3\phi\cdot z\\
  &\quad
  +16(a-2)(a-1)\bigl(45a^2-39a+6+4(15a^2-17a+3)\cos 2\phi\\
  &\qquad
  +(3a-4)(5a-3)\cos 4\phi\bigr)\sin^4\phi\Bigr).
\end{align*}
-----------------------}

We have also obtained 9-term recurrence relations for
$\mathcal{D}=\{1,3\},\{1,2,3\}$ with $X(\eta)=X_{\text{min}}(\eta)$.
Since the explicit forms of $r_{n,k}^{X,\mathcal{D}}$ are somewhat lengthy,
we do not write down them here.

\subsection{Examples for \S\,\ref{sec:ex_cH}}
\label{app:ex_cH}

We set $\sigma_1=a_1+a_1^*$, $\sigma_2=a_1a_1^*$,
$\sigma'_1=a_2+a_2^*$ and $\sigma'_2=a_2a_2^*$.

\noindent
\underline{Ex.1} $\mathcal{D}=\{1^{\I}\}$ ($M_{\I}=1,\,M_{\II}=0$), $Y(\eta)=1$
($\Rightarrow\ell_{\mathcal{D}}=1,\,X(\eta)=X_{\text{min}}(\eta),\,L=2$):
5-term recurrence relations
\begin{align*}
  X(\eta)&=\frac{\eta}{2}\bigl(
  (2-\sigma_1+\sigma'_1)\eta-2i(a_2-a_2^*+a_1a_2^*-a_1^*a_2)\bigr),\\
  r_{n,2}^{X,\mathcal{D}}&=
  \frac{(2-\sigma_1+\sigma'_1)(n+1)_2\,(n+\sigma_1-2)(n+b_1-1)_2}
  {2(n+\sigma_1)(2n+b_1-1)_4},\\
  r_{n,1}^{X,\mathcal{D}}&=
  \frac{-i(a_1-a_1^*-a_2+a_2^*)
  (b_1-2)(n+1)(n+\sigma_1-2)(n+\sigma'_1+1)(n+b_1-1)}
  {(2n+b_1-2)_3(2n+b_1+2)}.
\end{align*}
The polynomial $I(z)$ \eqref{Iz} is
\begin{align*}
  &\quad I(z)\\
  &=-4(2-\sigma_1+\sigma'_1)z^3\\
  &\quad
  +2\bigl(
  16-\sigma_1^3-12\sigma_2+12a_2\sigma_2+6a_1^2(2a_2-\sigma'_1)+14\sigma'_1
  +4\sigma_2\sigma'_1-3\sigma_1^{\prime\,2}
  +\sigma_1^{\prime\,3}\\
  &\qquad
  +\sigma_1^2(3-6a_2+3\sigma'_1)
  +\sigma_1(-10+12a_2+6a_2^2+6\sigma_2-16\sigma'_1-3\sigma_1^{\prime\,2}
  -4\sigma'_2)+20\sigma'_2\\
  &\qquad
  -6\sigma'_1\sigma'_2
  -6a_1(4a_2+2a_2^2-2\sigma'_1-\sigma_1^{\prime\,2}+2\sigma'_2)
  \bigr)z^2\\
  &\quad
  -2\Bigl(
  6-24\sigma_2+48a_2\sigma_2+16a_2^2\sigma_2+4a_2^3\sigma_2
  +4a_2^2a_2^*\sigma_2+27\sigma'_1-20\sigma_2\sigma'_1
  +4\sigma_1^{\prime\,2}+10\sigma_2\sigma_1^{\prime\,2}\\
  &\qquad
  -11\sigma_1^{\prime\,3}-10\sigma_2\sigma_1^{\prime\,3}+3\sigma_1^{\prime\,4}
  +\sigma_1^4(1+2\sigma'_1)+16\sigma_2\sigma'_2+24\sigma'_1\sigma'_2
  +4\sigma_2\sigma'_1\sigma'_2-14\sigma_1^{\prime\,2}\sigma'_2\\
  &\qquad
  -2a_1^3(6a_2+2a_2^2-3\sigma'_1-\sigma_1^{\prime\,2}+2\sigma'_2)
  +\sigma_1^3(-1+6a_2+2a_2^2-8\sigma'_1+10\sigma'_2)\\
  &\qquad
  -\sigma_1^2\bigl(8+24a_2+2a_2^3+2a_2^2(4+a_2^*)-19\sigma'_1
  -12\sigma_1^{\prime\,2}
  +2\sigma_2(3+5\sigma'_1)+46\sigma'_2-12\sigma'_1\sigma'_2\bigr)\\
  &\qquad
  +2a_1\bigl(-24a_2+2a_2^3+2a_2^2(-4+a_2^*)+4\sigma_1^{\prime\,2}
  -\sigma_1^{\prime\,3}-8\sigma'_2+2\sigma'_1(6+\sigma'_2)\bigr)\\
  &\qquad
  +2a_1^2\bigl(-6(-4+a_1^*)a_2+2a_2^3+a_2^2(8-2a_1^*+2a_2^*)
  +(-4+a_1^*)\sigma_1^{\prime\,2}-\sigma_1^{\prime\,3}\\
  &\qquad\quad
  -2(-4+a_1^*)\sigma'_2
  +\sigma'_1(-12+3a_1^*+2\sigma'_2)\bigr)\\
  &\qquad
  -\sigma_1\bigl(-9+2a_2^3+12a_2(-2+\sigma_2)
  +2a_2^2(-4+a_2^*+2\sigma_2)+40\sigma'_1+23\sigma_1^{\prime\,2}
  -12\sigma_1^{\prime\,3}\\
  &\qquad\quad
  +2\sigma_1^{\prime\,4}-52\sigma'_2
  +42\sigma'_1\sigma'_2-10\sigma_1^{\prime\,2}\sigma'_2
  +2\sigma_2(-12-15\sigma'_1+6\sigma_1^{\prime\,2}+2\sigma'_2)\bigr)
  \Bigr)z\\
  &\quad
  -\frac12(b_1-3)_2\Bigl(
  -32a_2^2\sigma_2-8a_2^3\sigma_2-8a_2^2a_2^*\sigma_2
  -2\sigma'_1+16\sigma_2\sigma'_1-7\sigma_1^{\prime\,2}
  +20\sigma_2\sigma_1^{\prime\,2}-5\sigma_1^{\prime\,3}\\
  &\qquad
  +\sigma_1^3\bigl(-4a_2^2+(1+2\sigma'_1)^2\bigr)
  -4a_1^2\bigl(2a_2^3+a_2^2(8-2a_1^*+2a_2^*)
  +(-4+a_1^*-\sigma'_1)(\sigma_1^{\prime\,2}-2\sigma'_2)\bigr)\\
  &\qquad
  +8a_1\bigl(2a_2^3+2a_2^2(2+a_2^*)-(2+\sigma'_1)(\sigma_1^{\prime\,2}
  -2\sigma'_2)\bigr)
  -32\sigma_2\sigma'_2+8\sigma'_1\sigma'_2-8\sigma_2\sigma'_1\sigma'_2\\
  &\qquad
  +a_1^3(8a_2^2-4\sigma_1^{\prime\,2}+8\sigma'_2)
  +\sigma_1^2\bigl(-1+4a_2^3+4a_2^2(4+a_2^*)+\sigma'_1-8\sigma_1^{\prime\,2}
  -4\sigma_1^{\prime\,3}+12\sigma'_2+16\sigma'_1\sigma'_2\bigr)\\
  &\qquad
  -\sigma_1\bigl(2+8a_2^3+8a_2^2(2+a_2^*-\sigma_2)
  +(-11+16\sigma_2)\sigma_1^{\prime\,2}-12\sigma_1^{\prime\,3}
  +24\sigma'_2-8\sigma_2\sigma'_2\\
  &\qquad\quad
  +4\sigma'_1(2+5\sigma_2+9\sigma'_2)\bigr)
  \Bigr).
\end{align*}

The example with $\mathcal{D}=\{1^{\II}\}$ ($M_{\I}=0,\,M_{\II}=1$) and
$Y(\eta)=1$ can be obtained by exchanging $a_1$ and $a_2$.

We have also obtained 9-term recurrence relations for
$\mathcal{D}=\{3^{\I}\},\{3^{\II}\},\{1^{\I},1^{\II}\}$ with
$X(\eta)$ $=X_{\text{min}}(\eta)$.
Since the explicit forms of $r_{n,k}^{X,\mathcal{D}}$ are somewhat lengthy,
we do not write down them here.

\subsection{Explicit form of $A$ in \eqref{A_Ex1_cH}}
\label{app:AinEx1cH}

\begin{align}
  &\quad
  \text{$A$ in \eqref{A_Ex1_cH}}\n
  &=(\sigma_1-\sigma'_1-4)_2\,n^5
  +(\sigma_1-\sigma'_1-4)_2\,(1+2b_1)n^4\n
  &\quad
  +\Bigl(-36+12\sigma_2-38a_2\sigma_2+12a_2^2\sigma_2
  +5a_1^3(2a_2-\sigma'_1)-5\sigma'_1
  +6\sigma_2\sigma'_1+29\sigma^{\prime\,2}_1-\sigma_2\sigma^{\prime\,2}_1\n
  &\qquad
  +16\sigma^{\prime\,3}_1
  +\sigma_1^3(-2-5a_2+\sigma'_1)
  +\bigl(12+12\sigma_2+\sigma'_1(-13+5\sigma'_1)\bigr)\sigma'_2
  -\sigma_1^2\bigl(5+a_2(-19+6a_2)\n
  &\qquad
  -5\sigma_2+\sigma'_1(19+2\sigma'_1)+\sigma'_2\bigr)
  +a_1^2\bigl(2a_2(-19+5a_1^*+6a_2)+19\sigma'_1-5a_1^*\sigma'_1
  -6\sigma^{\prime\,2}_1+12\sigma'_2\bigr)\n
  &\qquad
  +\sigma_1\bigl(45-5a_2^3+a_2^2(13-5a_2^*)+2a_2(-6+5\sigma_2)
  +\sigma_2(-19+\sigma'_1)
  -6\sigma'_2+\sigma'_1(52\n
  &\qquad
  +\sigma'_1(5+\sigma'_1)+\sigma'_2)\bigr)
  +a_1\bigl(2a_2(12+a_2(-13+5\sigma'_1))-26\sigma'_2
  +\sigma'_1(-12+13\sigma'_1-5\sigma^{\prime\,2}_1\n
  &\qquad
  +10\sigma'_2)\bigr)\Bigr)n^3\n
  &\quad
  +\Bigl(-36+5a_2^4(2a_1-\sigma_1)+10a_2^3a_2^*(2a_1-\sigma_1)
  -15\sigma_1+30\sigma_1^2-7\sigma_1^3+3\sigma_1^4-\sigma_1^5+12\sigma_2\n
  &\qquad
  -7\sigma_1\sigma_2-14\sigma_1^2\sigma_2+5\sigma_1^3\sigma_2
  +a_2^3\bigl(2a_1(-8+11a_1)+(8-11\sigma_1)\sigma_1+22\sigma_2\bigr)\n
  &\qquad
  +a_2^2a_2^*\bigl(2a_1(-8+11a_1)+(8-11\sigma_1)\sigma_1+22\sigma_2\bigr)
  +a_2\bigl(2a_1(12+a_1(-7-14a_1+5a_1^2\n
  &\qquad
  +2(-7+5a_1)a_1^*))-(-3+\sigma_1)\sigma_1(1+\sigma_1)(-4+5\sigma_1)
  +2(-7+\sigma_1(-14+5\sigma_1))\sigma_2\n
  &\qquad
  +10\sigma_2^2\bigr)+a_2^2\bigl(2a_1(-1+a_1(-26+11\sigma_1))-52\sigma_2
  +\sigma_1(1+(26-11\sigma_1)\sigma_1+22\sigma_2)\bigr)\n
  &\qquad
  +\bigl(-65+a_1(-12+a_1(7+a_1(14-5a_1-10a_1^*)+14a_1^*))+44\sigma_1
  +8\sigma_1^2-6\sigma_1^3\n
  &\qquad
  +6(-1+\sigma_1)^2\sigma_2-5\sigma_2^2\bigr)\sigma'_1
  +\bigl(a_1(1+a_1(26-11\sigma_1))+\sigma_1(38+(-12+\sigma_1)\sigma_1)\n
  &\qquad
  +2(-7+9\sigma_2)\bigr)\sigma^{\prime\,2}_1+\bigl(17+(8-11a_1)a_1
  +\sigma_1(8+\sigma_1)-6\sigma_2\bigr)\sigma^{\prime\,3}_1
  +(7-5a_1)\sigma^{\prime\,4}_1\n
  &\qquad
  -\sigma^{\prime\,5}_1-2\bigl(-6+a_1+a_1^2(26-11\sigma_1)
  +3(-1+\sigma_1)^2\sigma_1+26\sigma_2-11\sigma_1\sigma_2\bigr)\sigma'_2\n
  &\qquad
  +\bigl(-1+2a_1(-8+11a_1)-18\sigma_1+22\sigma_2\bigr)\sigma'_1\sigma'_2
  +2(-4+5a_1+3\sigma_1)\sigma^{\prime\,2}_1\sigma'_2
  +5\sigma^{\prime\,3}_1\sigma'_2\n
  &\qquad
  +5(2a_1-\sigma_1)\sigma^{\prime\,2}_2\Bigr)n^2\n
  &\quad
  +\Bigl(24-50\sigma_1+20\sigma_1^2-8\sigma_1^3+8\sigma_1^4
  -2\sigma_1^5-12\sigma_2+46\sigma_1\sigma_2
  -41\sigma_1^2\sigma_2+9\sigma_1^3\sigma_2\n
  &\qquad
  +a_2^4\bigl(6a_1+8a_1^2
  -\sigma_1(3+4\sigma_1)+8\sigma_2\bigr)
  +2a_2^3a_2^*\bigl(6a_1+8a_1^2-\sigma_1(3+4\sigma_1)+8\sigma_2\bigr)\n
  &\qquad
  +a_2\bigl(2a_1(-12+a_1(46-41a_1^*+a_1(-41+9a_1+18a_1^*)))
  -(-3+\sigma_1)\sigma_1(4+\sigma_1(-14\n
  &\qquad
  +9\sigma_1))
  +92\sigma_2+2\sigma_1(-41+9\sigma_1)\sigma_2
  +18\sigma_2^2\bigr)+a_2^2\bigl(2a_1(10+a_1(-22-3a_1^*+a_1(-3\n
  &\qquad
  +4a_1+8a_1^*)))
  +3\sigma_1^3-4\sigma_1^4+4\sigma_2(-11+2\sigma_2)-2\sigma_1(5+3\sigma_2)
  +\sigma_1^2(22+8\sigma_2)\bigr)\n
  &\qquad
  +a_2^2\sigma'_1\bigl(2a_1(-5+a_1(-9+8\sigma_1))-18\sigma_2
  +\sigma_1(5+(9-8\sigma_1)\sigma_1+16\sigma_2)\bigr)\n
  &\qquad
  -\bigl(14+a_1(-12+a_1(46-41a_1^*+a_1(-41+9a_1+18a_1^*)))
  +\sigma_1(40+\sigma_1(-46\n
  &\qquad
  +\sigma_1(18+(-5+\sigma_1)\sigma_1)))+36\sigma_2
  +\sigma_1(-19-4(-3+\sigma_1)\sigma_1)\sigma_2+9\sigma_2^2\bigr)\sigma'_1
  +\bigl(-36\n
  &\qquad
  +a_1(-10+a_1(22+a_1(3-4a_1-8a_1^*)+3a_1^*))-3\sigma_1^3
  +\sigma_1(14-6\sigma_2)+(17-4\sigma_2)\sigma_2\n
  &\qquad
  +\sigma_1^2(6+4\sigma_2)\bigr)\sigma^{\prime\,2}_1
  +\bigl(-4+a_1(5+a_1(9-8\sigma_1))
  +\sigma_1(14+\sigma_1(-5+2\sigma_1)-4\sigma_2)\n
  &\qquad
  +12\sigma_2\bigr)\sigma^{\prime\,3}_1+\bigl(6-a_1(3+4a_1)+5\sigma_1
  -4\sigma_2\bigr)\sigma^{\prime\,4}_1-\sigma_1\sigma^{\prime\,5}_1
  +\bigl(-12+2a_1(10+a_1(-22\n
  &\qquad
  -3a_1^*+a_1(-3+4a_1+8a_1^*)))+12\sigma_1^3
  -4\sigma_1^4-6\sigma_1(-6+\sigma_2)-44\sigma_2+8\sigma_2^2
  \n
  &\qquad
  +\sigma_1^2(-19+8\sigma_2)\bigr)\sigma'_2
  +\bigl(10+2a_1(-5+a_1(-9+8\sigma_1))-18\sigma_2
  +\sigma_1(-17+6\sigma_1-4\sigma_1^2\n
  &\qquad
  +16\sigma_2)\bigr)\sigma'_1\sigma'_2
  +\bigl(-5+6a_1+8a_1^2+4(-3+\sigma_1)\sigma_1
  +8\sigma_2\bigr)\sigma^{\prime\,2}_1\sigma'_2
  +(3+4\sigma_1)\sigma^{\prime\,3}_1\sigma'_2\n
  &\qquad
  +\bigl(6a_1+8a_1^2
  -\sigma_1(3+4\sigma_1)+8\sigma_2\bigr)\sigma^{\prime\,2}_2\Bigr)n\n
  &\quad
  +(1+\sigma'_1)(b_1-3)_2
  \Bigl(4+\sigma_1-2\sigma_2+2a_2(\sigma_1-2\sigma_1^2+4\sigma_2)
  +a_1^2(8a_2-4\sigma'_1)+3\sigma'_1\n
  &\qquad
  +2a_1(-2a_2+\sigma'_1)-(\sigma_1-\sigma'_1)(\sigma_1^2-4\sigma_2
  +\sigma'_1-\sigma_1\sigma'_1)-2\sigma'_2+ 4\sigma_1\sigma'_2\Bigr).
  \label{AinEx1cH}
\end{align}



\begin{thebibliography}{99}
%

\bibitem{kls}
R.\,Koekoek, P.\,A.\,Lesky and R.\,F.\,Swarttouw,
{\it Hypergeometric orthogonal polynomials and their $q$-analogues,\/}
Springer-Verlag Berlin-Heidelberg (2010).

\bibitem{os13}
S.\,Odake and R.\,Sasaki,
``Exactly solvable `discrete' quantum mechanics;
shape invariance, Heisenberg solutions,
annihilation-creation operators and coherent states,''
Prog. Theor. Phys. {\bf 119} (2008) 663-700.
{\tt arXiv:0802.1075[quant-ph]}.

\bibitem{os15}
S.\,Odake and R.\,Sasaki,
``Crum's theorem for `discrete' quantum mechanics,''
Prog. Theor. Phys. {\bf 122} (2009) 1067-1079,
{\tt arXiv:0902.2593[math-ph]}.

\bibitem{gos}
L.\,Garc\'ia-Guti\'errez, S.\,Odake and R.\,Sasaki,
``Modification of Crum's theorem for `discrete' quantum mechanics,''
Prog. Theor. Phys. {\bf 124} (2010) 1-26,
{\tt arXiv:1004.0289\hspace{0pt}[math-ph]}.

\bibitem{os24}
S.\,Odake and R.\,Sasaki,
``Discrete quantum mechanics,'' (Topical Review)
J. Phys. {\bf A44} (2011) 353001 (47pp),
{\tt arXiv:1104.0473[math-ph]}.

\bibitem{os12}
S.\,Odake and R.\,Sasaki,
``Orthogonal Polynomials from Hermitian Matrices,''
J. Math. Phys. {\bf 49} (2008) 053503 (43pp),
{\tt arXiv:0712.4106[math.CA]}.

\bibitem{os22}
S.\,Odake and R.\,Sasaki,
``Dual Christoffel transformations,''
Prog. Theor. Phys. {\bf 126} (2011) 1-34,
{\tt arXiv:1101.5468[math-ph]}.

\bibitem{os34}
S.\,Odake and R.\,Sasaki,
``Orthogonal Polynomials from Hermitian Matrices $\II$,''
J. Math. Phys. {\bf 59} (2018) 013504 (42pp),
{\tt arXiv:1604.00714[math.CA]}.

\bibitem{gkm08}
D.\,G\'{o}mez-Ullate, N.\,Kamran and R.\,Milson,
``An extension of Bochner's problem: exceptional invariant subspaces,''
J. Approx. Theory {\bf 162} (2010) 987-1006,
{\tt arXiv:\hspace{0pt}0805.\hspace{0pt}3376[math-ph]};
%
``An extended class of orthogonal polynomials defined by a
Sturm-Liouville problem,''
J. Math. Anal. Appl. {\bf 359} (2009) 352-367,
{\tt arXiv:0807.3939[math-\hspace{0pt}ph]}.

\bibitem{q08}
C.\,Quesne,
``Exceptional orthogonal polynomials, exactly solvable potentials
and supersymmetry,''
J. Phys. {\bf A41} (2008) 392001 (6pp),
{\tt arXiv:0807.4087[quant-ph]}.

\bibitem{os16}
S.\,Odake and R.\,Sasaki,
``Infinitely many shape invariant potentials and new orthogonal polynomials,''
Phys. Lett. {\bf B679} (2009) 414-417,
{\tt arXiv:0906.0142[math-ph]}.

\bibitem{os19}
S.\,Odake and R.\,Sasaki,
``Another set of infinitely many exceptional ($X_{\ell}$) Laguerre
polynomials,''
Phys. Lett. {\bf B684} (2010) 173-176,
{\tt arXiv:0911.3442[math-ph]}.

\bibitem{gkm11} 
D.\,G\'{o}mez-Ullate, N.\,Kamran and R.\,Milson,
``On orthogonal polynomials spanning a non-standard flag,''
Contemp. Math. {\bf 563} (2011) 51-72,
{\tt arXiv:1101.5584[math-ph]}.

\bibitem{gkm11_2} 
D.\,G\'{o}mez-Ullate, N.\,Kamran and R.\,Milson,
``Two-step Darboux transformations and exceptional Laguerre polynomials,''
J. Math. Anal. Appl. {\bf 387} (2012) 410-418,
{\tt arXiv:\hspace{0pt}1103.5724[math-ph]}.

\bibitem{os25}
S.\,Odake and R.\,Sasaki,
``Exactly solvable quantum mechanics and infinite families of
multi-indexed orthogonal polynomials,''
Phys. Lett. {\bf B702} (2011) 164-170,
{\tt arXiv:1105.\hspace{0pt}0508[math-ph]}.

\bibitem{os17}
S.\,Odake and R.\,Sasaki,
``Infinitely many shape invariant discrete quantum mechanical systems
and new exceptional orthogonal polynomials related to the Wilson and
Askey-Wilson polynomials,''
Phys. Lett. {\bf B682} (2009) 130-136,
{\tt arXiv:0909.3668[math-ph]}.

\bibitem{os27}
S.\,Odake and R.\,Sasaki,
``Multi-indexed Wilson and Askey-Wilson polynomials,''
J. Phys. {\bf A46} (2013) 045204 (22pp),
{\tt arXiv:1207.5584[math-ph]}.

\bibitem{os23}
S.\,Odake and R.\,Sasaki,
``Exceptional ($X_{\ell}$) ($q$)-Racah polynomials,''
Prog. Theor. Phys. {\bf 125} (2011) 851-870,
{\tt arXiv:1102.0812[math-ph]}.

\bibitem{os26}
S.\,Odake and R.\,Sasaki,
``Multi-indexed ($q$-)Racah polynomials,''
J. Phys. {\bf A 45} (2012) 385201 (21pp),
{\tt arXiv:1203.5868[math-ph]}.

\bibitem{ggm13}
D.\,G\'{o}mez-Ullate, Y.\,Grandati and R.\,Milson,
``Rational extensions of the quantum harmonic oscillator and exceptional
Hermite polynomials,''
J. Phys. {\bf A47} (2014) 015203 (27pp),
{\tt arXiv:1306.5143[math-ph]}.

\bibitem{d13}
A.\,J.\,Dur\'{a}n,
``Exceptional Meixner and Laguerre orthogonal polynomials,''
J. Approx. Theory {\bf 184} (2014) 176-208,
{\tt arXiv:1310.4658[math.CA]}.

\bibitem{os35}
S.\,Odake and R.\,Sasaki,
``Multi-indexed Meixner and Little $q$-Jacobi (Laguerre) Polynomials,''
J. Phys. {\bf A50} (2017) 165204 (23pp),
{\tt arXiv:1610.09854[math.CA]}.

\bibitem{detmiop}
S.\,Odake,
``New Determinant Expressions of the Multi-indexed Orthogonal Polynomials
in Discrete Quantum Mechanics,''
Prog. Theor. Exp. Phy. {\bf 2017(5)} (2017) 053A01 (36pp), 
{\tt arXiv:1702.03078[math-ph]}.

\bibitem{dualmiopqR}
S.\,Odake,
``Dual Polynomials of the Multi-Indexed ($q$-)Racah Orthogonal Polynomials,''
Prog. Theor. Exp. Phy. {\bf 2018} (2018) 073A02 (23pp),
{\tt arXiv:1805.00345[math-ph]}.

\bibitem{bochner}
E.\,Routh,
``On some properties of certain solutions of a differential equation
of the second order,''
Proc. London Math. Soc. {\bf 16} (1884) 245-261;
%
S.\,Bochner,
``\"Uber Sturm-Liouvillesche Polynomsysteme,''
Math. Zeit. {\bf 29} (1929) 730-736.

\bibitem{szego}
G.\,Szeg\"o,
{\it Orthogonal polynomials},
Amer. Math. Soc., Providence, RI (1939);
%
T.\,S. Chihara,
{\it An Introduction to Orthogonal Polynomials},
Gordon and Breach, New York (1978);
%
M.\,E.\,H.\,Ismail,
{\it Classical and Quantum Orthogonal Polynomials in One Variable\/},
vol. 98 of Encyclopedia of mathematics and its applications,
Cambridge Univ. Press, Cambridge (2005).

\bibitem{stz10}
R.\,Sasaki, S.\,Tsujimoto and A.\,Zhedanov,
``Exceptional Laguerre and Jacobi polynomials and the corresponding
potentials through Darboux-Crum transformations,''
J. Phys. {\bf A43} (2010) 315204,
{\tt arXiv:1004.4711[math-ph]}.

\bibitem{rrmiop}
S.\,Odake,
``Recurrence Relations of the Multi-Indexed Orthogonal Polynomials,''
J. Math. Phys. {\bf 54} (2013) 083506 (18pp),
{\tt arXiv:1303.5820[math-ph]}.

\bibitem{d14}
A.\,J.\,Dur\'{a}n,
``Higher order recurrence relation for exceptional Charlier, Meixner,
Hermite and Laguerre orthogonal polynomials,''
Integral Transforms Spec. Funct. {\bf 26} (2015) 357-376,
{\tt arXiv:1409.4697[math.CA]}.

\bibitem{mt14}
H.\,Miki and S.\,Tsujimoto,
``A new recurrence formula for generic exceptional orthogonal polynomials,''
J. Math. Phys. {\bf 56} (2015) 033502 (13pp),
{\tt arXiv:1410.0183[math.CA]}.

\bibitem{rrmiop2}
S.\,Odake,
``Recurrence Relations of the Multi-Indexed Orthogonal Polynomials : $\II$,''
J. Math. Phys. {\bf 56} (2015) 053506 (18pp),
{\tt arXiv:1410.8236[math-ph]}.
%

\bibitem{gkkm15}
D.\,G\'{o}mez-Ullate, A.\,Kasman, A.\,B.\,J.\,Kuijlaars and R.\,Milson,
``Recurrence Relations for Exceptional Hermite Polynomials,''
J. Approx. Theory {\bf 204} (2016) 1-16,
{\tt arXiv:\hspace{0pt}1506.03651[math.CA]}.

\bibitem{rrmiop3}
S.\,Odake,
``Recurrence Relations of the Multi-Indexed Orthogonal Polynomials : $\III$,''
J. Math. Phys. {\bf 57} (2016) 023514 (24pp),
{\tt arXiv:1509.08213[math-ph]}.

\bibitem{rrmiop4}
S.\,Odake,
``Recurrence Relations of the Multi-Indexed Orthogonal Polynomials $\IV$ :
closure relations and creation/annihilation operators,''
J. Math. Phys. {\bf 57} (2016) 113503 (22pp),
{\tt arXiv:1606.02836[math-ph]}.

\bibitem{rrmiop5}
S.\,Odake,
``Recurrence Relations of the Multi-Indexed Orthogonal Polynomials $\V$ :
Racah and $q$-Racah types,''
J. Math. Phys. {\bf 60} (2019) 023508 (30pp),
{\tt arXiv:1804.10352\hspace{0mm}[math-ph]}.

\bibitem{idQMcH}
S.\,Odake,
``Exactly Solvable Discrete Quantum Mechanical Systems and Multi-indexed
Orthogonal Polynomials of the Continuous Hahn and Meixner-Pollaczek Types,''
Prog. Theor. Exp. Phy. {\bf 2019} (2019) 123A01 (20pp),
{\tt arXiv:1907.12218[math-ph]}.

\bibitem{os29}
S.\,Odake and R.\,Sasaki,
``Krein-Adler transformations for shape-invariant potentials and pseudo
virtual states,"
J. Phys. {\bf A46} (2013) 245201 (24pp),
{\tt arXiv:1212.6595[math-\hspace{0pt}ph]}.

\bibitem{os30}
S.\,Odake and R.\,Sasaki,
``Casoratian Identities for the Wilson and Askey-Wilson Polynomials,"
J. Approx. Theory {\bf 193} (2015) 184-209,
{\tt arXiv:1308.4240[math-ph]}.

\bibitem{casoidrdqm}
S.\,Odake,
``Casoratian Identities for the Discrete Orthogonal Polynomials
in Discrete Quantum Mechanics with Real Shifts,''
Prog. Theor. Exp. Phy. {\bf 2017(12)} (2017) 123A02 (30pp), 
{\tt arXiv:1708.01830[math-ph]}.

\bibitem{equiv_miop}
S.\,Odake,
``Equivalences of the Multi-Indexed Orthogonal Polynomials,"
J. Math. Phys. {\bf 55} (2014) 013502 (17pp),
{\tt arXiv:1309.2346[math-ph]}.

\bibitem{os14}
S.\,Odake and R.\,Sasaki,
``Unified theory of exactly and quasi-exactly solvable `discrete'
quantum mechanics: I. Formalism,''
J. Math. Phys {\bf 51} (2010) 083502 (24pp),
{\tt arXiv:\hspace{0pt}0903.2604[math-ph]}.

\bibitem{os32}
S.\,Odake and R.\,Sasaki,
``Solvable Discrete Quantum Mechanics: $q$-Orthogonal Polynomials
with $|q|=1$ and Quantum Dilogarithm,''
J. Math. Phys. {\bf 56} (2015) 073502 (25pp),
{\tt arXiv:1406.2768[math-ph]}.

\bibitem{os7}
S.\,Odake and R.\,Sasaki,
``Unified theory of annihilation-creation operators for solvable
(`discrete') quantum mechanics,''
J. Math. Phys. {\bf 47} (2006) 102102 (33pp),
{\tt arXiv:\hspace{0pt}quant-ph/0605215};
%
``Exact solution in the Heisenberg picture and annihilation-creation
operators,''
Phys. Lett. {\bf B641} (2006) 112-117,
{\tt arXiv:quant-ph/0605221}.

\end{thebibliography}
\end{document}